\documentclass[refereee,structabstract]{aa}
\usepackage[utf8]{inputenc}
\usepackage{txfonts}
\usepackage{graphicx}
\usepackage{natbib}
\usepackage{multirow}
\usepackage{upgreek}

\usepackage{amsmath}
\usepackage{url}
\usepackage{hyperref}
\hypersetup{
        bookmarksnumbered = true,
        colorlinks = true,
        unicode = true,
        urlcolor = blue,
        linkcolor = blue,
        citecolor = blue
        }

\bibpunct{(}{)}{;}{a}{}{,}

\DeclareMathAlphabet{\mathitsf}{\encodingdefault}{\sfdefault}{m}{sl}

\newcommand{\emm}[1]{\ensuremath{#1}}
\newcommand{\emr}[1]{\emm{\mathrm{#1}}}

\newcommand{\unit}[1]{\emr{\,#1}}
\newcommand{\e}[1]{\emm{\times 10^{#1}}}

\newcommand{\Av}{\emm{A_\emr{V}}}
\newcommand{\Ak}{\emm{A_\emr{K}}}

\newcommand{\kHz}{\unit{kHz}}

\newcommand{\GHz}{\unit{GHz}}

\newcommand{\Msun}{\unit{M_{\sun}} }
\newcommand{\pc}{\unit{pc}}
\newcommand{\mpc}{\unit{mpc}}
\newcommand{\pscm}{\unit{cm^{-2}}} % per square centimeter
\newcommand{\pccm}{\unit{cm^{-3}}} % per cubic centimeter

\newcommand{\micron}{\unit{\upmu m}}
\newcommand{\K}{\unit{K}}
\newcommand{\mm}{\unit{mm}}
\newcommand{\kms}{\unit{km\,s^{-1}}}

\newcommand{\Hii}{\textsc{Hii}}
\newcommand{\CO}{\emr{C}\emm{^{18}}\emr{O}}
\newcommand{\COline}{\CO{}\emm{\,(J=1-0)}}
\newcommand{\SNR}{signal-to-noise ratio}
\newcommand{\Nh}{\emm{N_\emr{H}}}

\begin{document}

\title{A dynamically young, gravitationally stable network of filaments in Orion\,B\thanks{Based on observations carried out at the IRAM 30\,m single-dish telescope. IRAM is supported by INSU/CNRS (France), MPG (Germany) and IGN (Spain).}}
%\titlerunning{Turbulence and star formation efficiency in Orion\,B}

\author{Jan H. Orkisz \inst{\ref{z:UGA},\ref{z:IRAM},\ref{z:LERMA-ENS}} %
  \and Nicolas Peretto \inst{\ref{z:UC}} %
  \and Jérôme Pety \inst{\ref{z:IRAM},\ref{z:LERMA-ENS}} %
  \and Maryvonne Gerin \inst{\ref{z:LERMA-ENS}} %
  \and François Levrier \inst{\ref{z:LERMA-ENS}} %
  \and Emeric Bron \inst{\ref{z:LERMA-M},\ref{z:CSIC}} %
  \and Sébastien Bardeau \inst{\ref{z:IRAM}} %
  \and Javier R. Goicoechea \inst{\ref{z:CSIC}} %
  \and Pierre Gratier \inst{\ref{z:LAB}} %
  \and Viviana V. Guzm\'an \inst{\ref{z:PUC}} %
  \and Annie Hughes \inst{\ref{z:IRAP}} %
  \and David Languignon \inst{\ref{z:LERMA-M}} %
  \and Franck Le Petit \inst{\ref{z:LERMA-M}} %
  \and Harvey S. Liszt \inst{\ref{z:NRAO}} %
  \and Karin \"Oberg \inst{\ref{z:CFA}} %
  \and Evelyne Roueff \inst{\ref{z:LERMA-M}} %
  \and Albrecht Sievers \inst{\ref{z:IRAM-SPAIN}} %
  \and Pascal Tremblin \inst{\ref{z:CEA}} %
}

\institute{%
  Univ. Grenoble Alpes, IRAM, F-38000 Grenoble, France \label{z:UGA} %
  \and IRAM, 300 rue de la Piscine, F-38406 Saint Martin d'Hères,
  France \label{z:IRAM} %
  \and Sorbonne Université, Observatoire de Paris, Université PSL, École normale supérieure, CNRS, LERMA, F-75005, Paris, France \label{z:LERMA-ENS} %
  \and School of Physics and Astronomy, Cardiff University, Queen's
  buildings, Cardiff CF24 3AA, UK \label{z:UC} %
  \and Sorbonne Université, Observatoire de Paris, Université PSL, CNRS, LERMA, F-92190, Meudon,
  France \label{z:LERMA-M} %
  \and Grupo de Astrofisica Molecular. IFF-CSIC. Calle Serrano 121-123. 28006, Madrid, Spain. \label{z:CSIC} %
  \and Laboratoire d'astrophysique de Bordeaux, Univ. Bordeaux, CNRS, B18N,
  allée Geoffroy Saint-Hilaire, 33615 Pessac, France \label{z:LAB} %
  \and Instituto de Astrof{\'i}sica, Ponticia Universidad Cat{\'o}lica de Chile, Av.~Vicu{\~n}a Mackenna 4860, 7820436 Macul, Santiago, Chile \label{z:PUC} %
  \and CNRS, IRAP, 9 Av. colonel Roche, BP 44346, 31028 Toulouse
Cedex 4, France \label{z:IRAP} %
  \and National Radio Astronomy Observatory, 520 Edgemont Road,
  Charlottesville, VA, 22903, USA \label{z:NRAO} %
  \and Harvard-Smithsonian Center for Astrophysics, 60 Garden Street,
  Cambridge, MA, 02138, USA \label{z:CFA} %
  \and IRAM, Avenida Divina Pastora, 7, Núcleo Central, E-18012 Granada,
  España \label{z:IRAM-SPAIN}%
  \and Maison de la Simulation, CEA-CNRS-INRIA-UPS-UVSQ, USR 3441, Centre
  d'étude de Saclay, F-91191 Gif-Sur-Yvette, France \label{z:CEA} %
}

\date{}

\abstract%
%{Filaments are a key step on the path that leads from molecular clouds to star formation in dense cores. Their study mostly focuses on a limited number of well-defined and easily identified structures. Their characteristics (for instance their width) are heavily debated, and their formation and fragmentation processes still need to be understood.}
{Filaments are a key step on the path that leads from molecular clouds to star formation. However, their characteristics, for instance their width, are heavily debated and the exact processes that lead to their formation and fragmentation into dense cores still remain to be fully understood.}%
{We aim at characterising the mass, kinematics, and stability against gravitational collapse of a statistically significant sample of filaments in the Orion\,B molecular cloud, which is renown for its very low star formation efficiency.}%
{We characterised the gas column densities and kinematics over a field of 1.9 deg$^2$, using \COline\ data from the IRAM 30\,m large programme ORION-B at angular and spectral resolutions of $23.5''$ and $49.5\,$kHz, respectively. Using two different Hessian-based filters, we extracted and compared two filamentary networks, each containing over 100 filaments.}%
{Independent of the extraction method, the filament networks have consistent characteristics. The filaments have widths of $\sim0.12\pm0.04\pc$ and show a wide range of linear ($\sim 1 - 100 \Msun\pc^{-1}$) and volume densities ($\sim2\e{3} - 2\e{5} \pccm$). Compared to previous studies, the filament population is dominated by low-density, thermally sub-critical structures, suggesting that most of the identified filaments are not collapsing to form stars. In fact, only $\sim 1\%$ of the Orion\,B cloud mass covered by our observations can be found in super-critical, star-forming filaments, explaining the low star formation efficiency of the region. The velocity profiles observed across the filaments show quiescence in the centre and coherency in the plane of the sky, even though these profiles are mostly supersonic.}%
{The filaments in Orion\,B apparently belong to a continuum which contains a few elements comparable to already studied star-forming filaments, for example in the IC\,5146, Aquila or Taurus regions, as well as many lower density, gravitationally unbound structures. This comprehensive study of the Orion\,B filaments shows that the mass fraction in super-critical filaments is a key factor in determining star formation efficiency.}


\maketitle{}

\section{Introduction}
 
For decades, filaments of interstellar dust and molecular gas have been known to represent an important structural element of star-forming regions in the Galaxy \citep[e.g.][]{schneider79}. The possible mechanisms leading to the formation of interstellar filaments are numerous, and can involve turbulence, gravity, magnetic fields, or any combination of these \citep[e.g.][]{padoan01,burkert04,hennebelle13,smith14,federrath16}. The presence of filaments in non-self-gravitating clouds \citep[e.g.][]{wardthompson10} suggests that turbulence plays a major role in filament formation. It has been proposed that filaments result from the intersection of sheet-like shock structures in supersonic turbulence \citep{pety00,padoan01}. Recently, \citet{hennebelle13} showed that in fact if compression is necessary to accumulate gas in the first place, shear is the main driver behind clump elongation, and magnetic fields help confine filamentary structures and therefore make them more long-lived. Compression from winds of OB associations are also believed to have formed some of the Pipe Nebula filaments \citep{peretto12}.  External ram pressure is not the only process leading to the formation of filaments; in particular in self-gravitating gas, self-gravity also has the effect of enhancing density anisotropies, and thus clump elongation \citep[e.g.][]{hartmann07, peretto07}. \citet{nagai98} showed how gas sheets in hydrostatic equilibrium threaded by a magnetic field can fragment into filaments that are parallel or perpendicular to the field lines, depending on the gas density. 

Lately, far-infrared and submillimetre observations of the sky made with the \emph{Herschel} space observatory revealed the tight link between the presence and properties of interstellar filaments and their ability to form stars \citep[e.g.][]{andre10,molinari10}. It has been shown that more than 70\% of gravitationally bound cores lie within thermally super-critical filaments, where the linear density $M_\mathrm{l}$ is larger than a critical value $M_\mathrm{l}^\mathrm{crit}$ above which the filaments become gravitationally unstable \citep{polychroni13, konyves15}. This suggests that most star-forming cores form as the result of gravitational instabilities occurring within unstable filaments \citep{andre14}. Another key proposition that has emerged from \emph{Herschel} observations of star-forming clouds in the Gould Belt is the potential universality of the width of interstellar filaments at $\sim0.1$~pc \citep{arzoumanian11}. This width seems to be relatively independent of the central column density of the filaments, which is surprising as the densest filaments would be expected to collapse quickly and radially into thin spindles. \citet{arzoumanian13} proposed that accretion onto super-critical filaments could maintain a constant filament diameter during the contraction of the filament. However, using velocity-resolved maps of the filament gas emission, \citet{hacar13,hacar18} proposed that the filament width is actually not universal; but the filament width depends on environment and has broader filaments in low-mass, star-forming regions and narrower filaments in massive star-forming regions. In this picture, all filaments have a linear density that is about critical, close to hydrostatic equilibrium, explaining why they would not collapse into spindles. The so-called universality of the filament width would then be an observational bias of dust continuum emission maps that would merge narrower, velocity coherent filaments \citep[called ``fibres'' by][]{hacar13} into one, non-velocity coherent elongated structure. Filaments formed in numerical simulations that involve the three key elements (turbulence, gravity, and magnetic fields) discussed above seem to agree with \emph{Herschel} results regarding the proposed universality of filament widths \citep[e.g.][]{kirk15,federrath16}, possibly linked to the scale at which the gas becomes subsonic, as first proposed by \citet{andre10}. However, the predicted decrease of the velocity dispersion towards the inner parts of filaments is not always observed \citep{williams18}. This question of the (non-)universality of filament widths is still very much debated \citep[e.g.][]{panopoulou17}.

Observationally, the statistical characterisation of filament kinematics is a difficult task because it requires a large amount of telescope time. Until recently, only studies dedicated to a few individual filaments have been performed \citep{busquet13,hacar13,kirk13,peretto13,peretto14,hacar16,williams18}. Only in the last few years, large surveys such as the Green Bank Ammonia Survey \citep[GAS;][]{friesen17} and the CARMA-NRO Orion Survey \citep{kong18} have started to observe Gould Belt clouds in molecular lines over several square degrees. Such large datasets are required to build statistically significant samples of filaments and characterise their dynamical properties. However, none of these large surveys have yet looked in detail at the filament population. We propose to use observations from the IRAM 30\,m large programme Outstanding Imaging of OrioN-B (ORION-B) to analyse the properties of the Orion\,B filaments over an area of 1.9 deg$^2$. This study will provide, for the first time, a complete picture of the filament population in the region and shed light on the origin of the low star formation efficiency (SFE) observed in Orion B.

This paper is organised as follows. In Section \ref{sec:observations}, we present the IRAM 30 m, \emph{Herschel}, and \emph{Planck} data used in this study, and the data processing to which the IRAM 30\,m position-position-velocity (PPV) cubes were submitted. Section \ref{sec:filaments} introduces the filamentary network in the molecular cloud, and briefly explains how it is detected; methodological details on the filament identification procedures can be found in Appendix \ref{app:method}. A statistical analysis of the physical properties that characterise these filaments is presented in Section \ref{sec:results} , while the implications of these properties for the structure and evolution of the filaments are discussed in Section \ref{sec:discussion}. Section \ref{sec:conclusion} summarises the results and provides outlooks for this work.

\section{Observational data and data processing}
\label{sec:observations}
We aim at studying the structure and dynamics of the filaments in the south-western part the Orion\,B cloud. To do this, we need to constrain the mass, temperature, kinematics, and magnetic field in this region. We use \emph{Planck} data to constrain the geometry of the source; we use the map of the \COline{} line observed as part of the ORION-B large programme, combined with a dust temperature map deduced from \emph{Herschel} Gould Belt Survey (HGBS) observations, to derive the other quantities. This section presents the datasets and steps applied to obtain an accurate map of the molecular hydrogen column density.

\subsection{Molecular lines from the IRAM 30\,m ORION-B large programme}
\label{sec:obs-iram}
The ORION-B project (Outstanding Radio-Imaging of OrioN B; co-PIs: Jérôme Pety and Maryvonne Gerin) aims at mapping about 5 square degrees of the southern part of the Orion\,B molecular cloud over most of the 3\mm{} band in about 850\,hours of IRAM 30\,m telescope time. The observing strategy and data reduction are discussed in detail in \citet{pety17}, and we recall just a few key numbers. The observations cover a bandwidth of 40\GHz{}, from 72 to 80 and from 84 to 116\GHz{}, in three tunings of the EMIR receivers. The angular resolution ranges from $36\arcsec$ to $22.5\arcsec$ for these frequencies. The median sensitivity of the Fast Fourier Transform Spectrometer (FTS) spectra ranges from 0.1\K{}  to 0.5\K{} in main beam temperature depending on the frequency at a spectral channel spacing of 195\kHz{}. The \emm{J=1-0} lines of the $^{13}$CO{} and \CO{} isotopologues are observed with a median noise of 0.12\K{} and a spatial resolution of $23\arcsec$. At the typical distance of $\sim 400$\pc{} of the Orion\,B cloud \citep{menten07,schlafly14,schaefer16}, this corresponds to a resolution of about 45\mpc. The higher spectral resolution but narrower bandwidth VESPA auto-correlator was used simultaneously to observe several lines, including \emm{^{13}}CO and \COline{}. The median noise for these data is 0.34\K\,$(T_\mathrm{mb})$ for a spectral channel spacing of 40\kHz{}. The PPV cubes of all identified molecular lines are all smoothed to a $29\arcsec$ (60\mpc) resolution by convolution with a Gaussian kernel, and resampled onto identical grids with a resolution of $9\arcsec\times 9\arcsec\times 0.5\kms$ (or 0.1\kms{} for VESPA). The spatial coordinates of the data cubes are centred onto the photodissociation region (PDR) of the Horsehead Nebula and rotated by $14\degr$ anti-clockwise with respect to RA-Dec \,(J2000) coordinates, so that the edge of the IC\,434 PDR, from which the famous Horsehead Nebula emerges, is vertical in the maps.

\begin{figure} % in the aa.cls, you should not use [!htb]
    \centering
    \includegraphics[width=\linewidth]{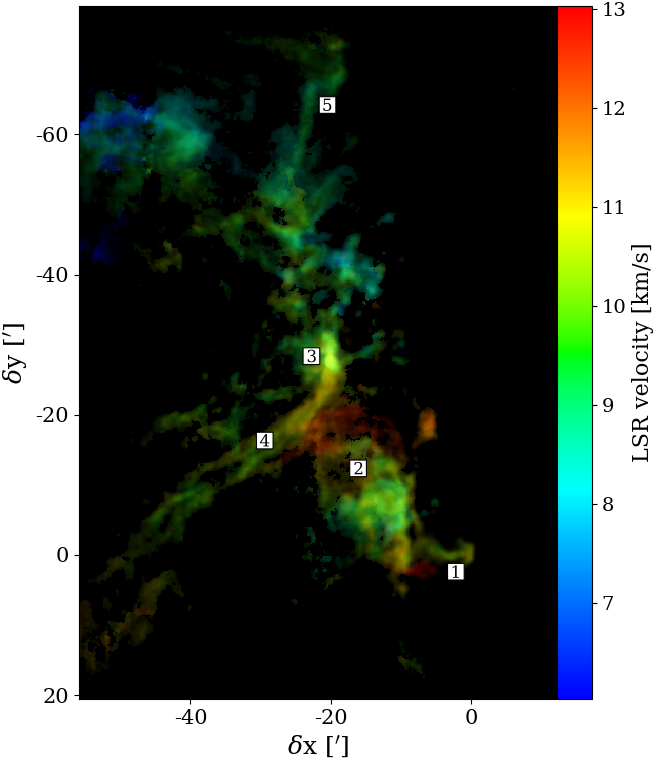}
    \caption{Composite image representing the velocity structure present in the \COline{} cube. Each velocity channel is directly encoded as a hue, and the final image is obtained by additive colour synthesis. The brightest points correspond to 4\,K in main beam temperature. The labels indicate (1) the Horsehead Nebula, (2) NGC\,2023, (3) NGC\,2024 (the Flame Nebula), (4) the Flame filament, and (5) the Hummingbird filament. The spatial and kinematic complexity of the cloud appear clearly in particular in regions such as the north-eastern part of the cloud or around NGC\,2023.}
    \label{fig:rgb}
  \end{figure}

As the data are still being acquired, the field presented in this paper covers 99\arcmin\ by 68\arcmin, which corresponds to about 11.5 by 7.9 parsecs at the distance of Orion\,B (Fig.~\ref{fig:rgb}).  It features objects such as the Horsehead pillar, the NGC\,2023 and NGC\,2024 (a.k.a. Flame Nebula) star-forming regions, and the PDRs IC\,434 (illuminated by the multiple system $\sigma$ Ori) and IC\,435 (illuminated by HD\,38087).

In this work we make use of the VESPA data because it provides a better spectral resolution while still retaining a good sensitivity, similar to the FTS sensitivity when smoothing the data to the same spectral resolution\footnote{The data products associated with this paper are available at \url{http://www.iram.fr/~pety/ORION-B}}. Most of the analysis focusses on the \COline{} data cube, which is our best available molecular tracer for the medium at high visual extinctions typical of filamentary structures. As \citet{pety17} have shown, the $^{12}$CO but also the $^{13}$CO$\,(J=1-0)$ lines are saturated at high visual extinctions, while the C$^{17}$O$\,(J=1-0)$ line is only detected around dense cores. In situations in which we compare the filaments with the more diffuse surrounding medium, we complement the \COline{} data with $^{13}$CO$\,(J=1-0)$.

Figure \ref{fig:rgb} shows a RGB representation of the complex velocity structure of the denser parts of Orion\,B, deduced from the  \COline{} PPV cube. A map of its peak temperature can also be seen in Fig. \ref{fig:filS1} (panel 1).

\subsection{Modelling the data cube as a sum of Gaussian profiles}
\label{sec:gauss}
The maximum value of the peak \SNR{} over the entire field of view in \COline{} is 63, while its median value is 2.8. It means that many lines of sight over the studied field of view are measured at a relatively low \SNR{}. The first step of our analysis is thus to transform the noisy observational data into a model cube. For that purpose, we performed a multi-Gaussian fit of the spectra for each individual line of sight. This provides a model data cube at a final spatial resolution of about 60\mpc\  (see Appendix \ref{app:gauss}).
The resulting model cube allows us to have a clean and easily exploitable representation of the noisy signal without the problematic windowing effects that may occur when \SNR{} masks are applied to the data.

\subsection{Independent column density and temperature estimate from the \emph{Herschel} Gould Belt Survey}
\label{sec:herschel}

By fitting a composite spectral energy distribution built from HGBS \citep[][]{andre10,schneider13} and \emph{Planck} satellite \citep{PCe29} continuum observations, \citet{lombardi14} derived a dust temperature map $T_\mathrm{dust}$ and a dust opacity map at 850\micron{}, $\tau_{850}$. These maps have a resolution of 36'', where HGBS coverage is available, and 5' otherwise. 

We converted the opacity map to a map of total hydrogen column density \Nh{} using the following conversion factors:
\begin{equation}
\begin{split}
\Nh{}/\Av{} &= 1.8\times 10^{21} \unit{cm}^{-2}\unit{mag}^{-1},\\
\Av{}/\tau_{850} &= 2.7 \times 10^4
\end{split}
.\end{equation}
These factors are based on \citet{bohlin78} and \citet{liszt14b}, and on \citet{rieke85} and \citet{cardelli89}, respectively \citep[see][for detailed explanations]{pety17}.

\begin{figure*}
    \centering
    \includegraphics[width=\linewidth]{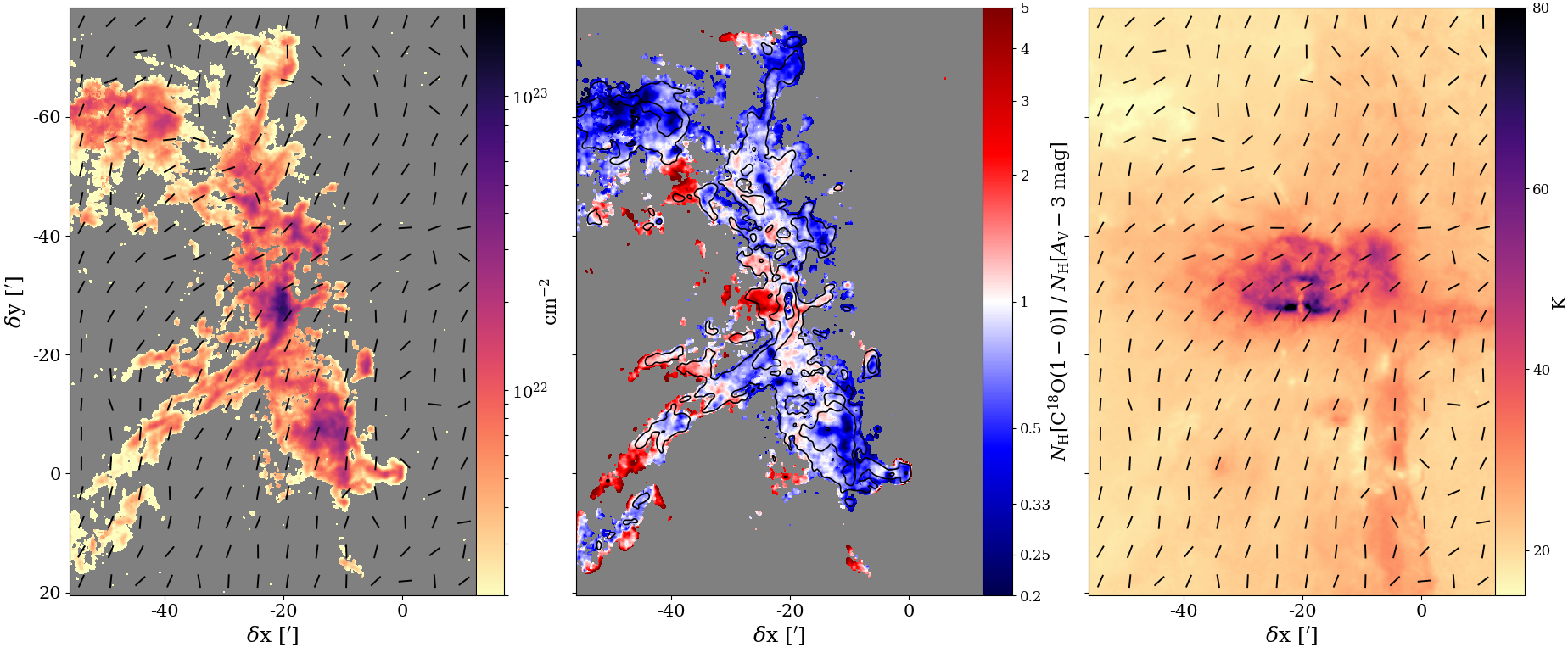}
    \caption{\emph{Left:} \Nh{} column density map derived from the \COline{} integrated intensity (multi-Gaussian model). The orientation of the magnetic field, derived from \emph{Planck} polarimetric data, is shown by the black segments. The spacing of the segments corresponds to the 5\arcmin\ beam of \emph{Planck}. \emph{Middle:} Map of the ratio of the \CO{}-traced \Nh{} to the dust-traced \Nh{} above 3 magnitudes of \Av\ in Orion\,B. Superimposed are contours of the \COline{} integrated intensity at 0.75, 2.5 and 7.5 \K{}\kms{}. \emph{Right:} Effective dust temperature map computed by \citet{lombardi14}. The black segments again show the orientation of the magnetic field.}
    \label{fig:data}
  \end{figure*}

The \COline{} line traces relatively dense $(n_\mathrm{H}\sim 10^3 - 10^4\pccm)$ material with low far-ultraviolet (FUV) illumination \citep{pety17,gratier17,bron18}. Although the dust-to-gas coupling is completely established only above $10^5\pccm$ \citep{goldsmith01}, the dust temperature should still be a better proxy for the gas temperature of this relatively cold medium than the excitation temperature of $^{12}$CO$\,(J=1-0)$, which traces the warmer, more diffuse envelope of the cloud \citep{bron18}. The dust temperature map is shown in Fig.~\ref{fig:data} (right). %This is a different procedure from the one used in \citet{orkisz17}, where we built a ``composite temperature'' map  using both the dust temperature and the $^{12}$CO$\,(J=1-0)$ peak temperature. This latter map was rather designed to represent the temperature of the diffuse and translucent medium where the density is too low for a complete coupling of the dust and gas temperatures, and thus better traces the temperature of the gas emitting the $^{13}$CO$\,(J=1-0)$ line.

\subsection{C$^\mathitsf{18}$O-derived column density}
\label{sec:c18o}
Using the integrated intensities from the modelled \COline{} cube, an estimate of the column density of \CO{} was computed assuming local thermodynamic equilibrium (LTE) and an optically thin medium in this line, and using the dust temperature from \citet{lombardi14}. We followed the standard equations described in \citet{mangum15}, using spectroscopic data from the Cologne Database for Molecular Spectroscopy (CDMS) \citep{muller05}. The \CO{} column density was converted to \Nh{} via $\Nh{} = 2\cdot N_{\CO}/5.6\times 10^{-7}$. This assumes that all the available carbon is locked in gas-phase CO and has a C/H$_2$ abundance of $\sim2.8\times10^{-4}$ \citep{sofia04,parvathi12,gerin15} and a $^{18}$O/$^{16}$O isotopic ratio of $\sim 1/500$ \citep{wilson94}.

%In addition to this \CO{}-based derivation of the \Nh{} column density, a $^{13}$CO-based derivation was performed as well for comparison. In this case, the composite temperature was used as a proxy for the gas temperature, as explained at the end of Sect.~\ref{sec:herschel}. The $^{13}$CO column density was converted to \Nh{} using $\Nh{} = 2\cdot N_\mathrm{^{13}CO}/4.7\times 10^{-6}$, which assumes a $^{13}$C/$^{12}$C isotopic ratio of $\sim 1/60$ \citep{wilson94,pineda13a} with negligible chemical fractionation and the same gas-phase carbon abundance.

\begin{figure}
    \centering
    \includegraphics[width=\linewidth]{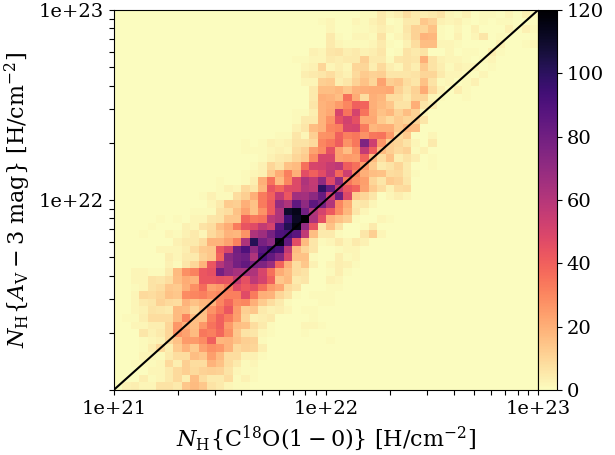}
    \caption{Joint distributions of the \Nh{} column densities as traced by \COline{} against those inferred from \Av{} with an offsets of 3 magnitudes. This threshold corresponds to the extinctions at which the molecular tracers starts to be detected according to \citet{pety17}. The distribution is computed in the identified filamentary regions. The 1:1 relation is overplotted as a black line.}
    %and $^{13}$CO$\,(J=1-0)$ against those inferred from \Av{} with offsets of 3 and 1 magnitude respectively. These thresholds correspond to the extinctions at which the molecular tracers start to be detected according to \citet{pety17}. The distributions are computed in the identified filamentary regions. The 1:1 relation is overplotted as a black line.}
    \label{fig:nhco}
  \end{figure}

The obtained column densities are not expected to trace the totality of the matter present along the line of sight, as there can be both atomic and molecular \CO{}-dark %(or $^{13}$CO-dark, respectively)
gas in the fore- or background of the molecular emission region. When comparing this molecular emission and the dust extinction \Av\ derived from \emph{Herschel} data, \citet{pety17} showed that the \CO\ % and $^{13}$CO 
emission starts to be detected at $\Av\sim3$. % and $\Av\sim1$, respectively.
We therefore compared the obtained \CO -traced column densities to the dust-traced column density above 1 magnitude of \Av, rather than to the total dust-traced column density. The comparison of these tracers in the filamentary regions identified and analysed in this work is shown in Fig.~\ref{fig:nhco}.
We see a good consistency of the resulting column densities. The ratio of $\Nh{}[\CO{}]/\Nh{}[\Av{}-3\unit{mag}]$ has a mean value of 1.02, a median of 0.83, and has a standard deviation of 1.19.

The middle panel of Fig.~\ref{fig:data} shows the spatial distribution of the $\Nh{}[\CO{}]/\Nh{}\left[\Av{}-3\unit{mag}\right]$ ratio, which is close to one in a large fraction of the map. It is significantly smaller than 1 on the western edge of the cloud, and in particular at the base of the Horsehead pillar. This might be an effect of selective photodissociation: at the edge of the IC\,434 PDR, the self-shielding of \CO{} is too weak to prevent its destruction by FUV radiation. The ratio is also much smaller than 1 in a region to the north-east, which is the coldest in the current field of view and is known to harbour dense cores. In that case, \CO{} depletion is the most probable explanation because the dust is cold enough for CO to freeze out on grain surfaces. The $\Nh{}[\CO{}]/\Nh{}[\Av{}-3\unit{mag}]$ ratio is conversely significantly larger than 1 in several low-density regions lying to the east. In these regions, away from the sources of photodissociating radiation, \Av{} ranges from 3.0 to 5.4, with an average of 3.5, close to the chosen extinction threshold of 3 magnitudes. In this region the column density ratio becomes very sensitive to the choice of the extinction threshold, leading to higher uncertainties. The extinction threshold may vary across the field of view and reach somewhat lower values in regions away from the interface with IC\,434. The ratio is also larger than 1 around NGC\,2024, which is mostly likely due to a layering effect with a significant temperature gradient along the line of sight, poorly rendered by a single value of effective dust temperature. As a consequence, the dust-traced column density is underestimated and the CO-traced column density is overestimated \citep{pety17}.

\begin{figure}
    \centering
    \includegraphics[width=\linewidth]{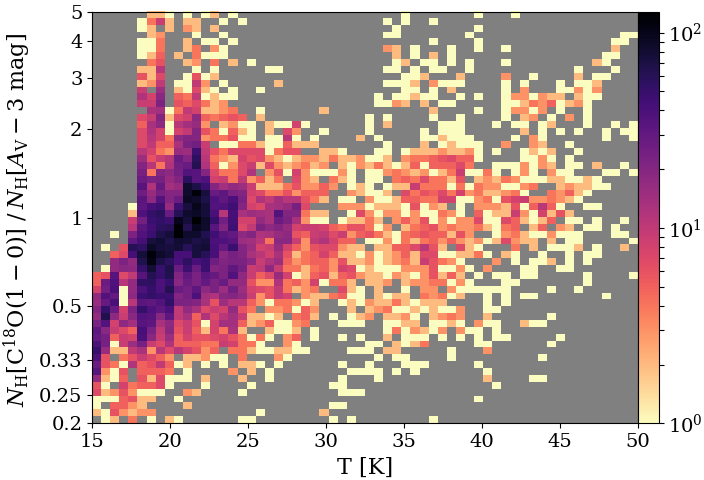}
    \caption{Joint distribution of the ratio of \CO{}-traced vs. dust-traced \Nh{} column density (Fig.~\ref{fig:data} middle) against the dust temperature (Fig.~\ref{fig:data} right), taken only in the identified filamentary regions.}
    \label{fig:depletion}
  \end{figure}

The question of depletion by freeze-out is important when studying filaments because they are expected to be dense and cold, and CO is known to freeze out at temperatures $< 20$\,K. We therefore compared the $\Nh{}[\CO{}]/\Nh{}[\Av{}-3]$ ratio to the temperature (Fig.~\ref{fig:depletion}) to search for potential signs of such a systematic depletion effect in the regions identified as filamentary (see Sect.~\ref{sec:extraction}). A tail, corresponding to both a lower temperature and a lower column density ratio, exists in the distribution. However, the lines of sight where CO is probably depleted (lines of sight with a temperature below 20\K{} and a column density ratio 50\% below its average value) amount to less than 8\% of the filamentary regions. Moreover, they almost exclusively lie in the north-eastern cold core region, which does not host major filaments (see Fig.~\ref{fig:rgb} and Fig.~\ref{fig:data}). For the most part, \COline\ is therefore a tracer well suited to recover the gas column densities in the density and temperature regimes of the filamentary regions of the Orion\,B cloud.

\subsection{Magnetic field orientation from the \emph{Planck} all-sky survey}
We also used \emph{Planck} polarisation data\footnote{Based on observations obtained with Planck (\url{http://www.esa.int/Planck}), an ESA science mission with instruments and contributions directly funded by ESA Member States, NASA, and Canada.} at 353\GHz{} to estimate the orientation of the magnetic field in Orion\,B. The Stokes $I$, $Q$, $U$ maps were used at their native resolution of 5' to derive the polarisation angle $\chi$ and the magnetic field angle $\psi$, which is rotated by 90\degr\ with respect to $\chi$ in the International Astronomical Union convention. These angles were rotated to match the custom north-south axis of our projection; i.e. 0\degr{} points to the top of the presented field, not to the standard north in J2000 equatorial coordinates.
The orientation of the magnetic field $\psi$ is presented in the left and right panels of Fig.~\ref{fig:data} superimposed on the column density map and temperature map, respectively.

\section{Detection and characterisation of the filamentary network}
\label{sec:filaments}
\subsection{Qualitative description of the filaments}
Figures \ref{fig:rgb} and \ref{fig:data} suggest that observations of \COline{} bring out a complex molecular filamentary network. At the very centre of the observed field, the star-forming region NGC\,2024 shines brightly. This spot has both the warmest temperature and highest column density in this area. The molecular emission comes from a dense filament, seen in the optical as a dark dusty lane in the foreground of the young \Hii{} region, with a characteristic shape that earned the Flame Nebula its name. This large Flame filament can be clearly seen in \COline{}, as it flows diagonally from NGC\,2024 to the south-east. In its more diffuse part, it is clearly sub-structured, made of parallel strands of molecular gas.

At the south-west of our field lies the Horsehead Nebula, with its characteristic shape. This nebula is a pillar carved in the Orion molecular cloud by the IC\,434 \Hii{} region.
Between NGC\,2024 and the Horsehead lies the quieter star-forming region NGC\,2023. The kinematics of the gas surrounding NGC\,2023 is complex; this region has at least two velocity components (visible in green and orange colours in Fig.~\ref{fig:rgb}), and the filaments this medium might host are less obviously distinguishable by eye. 

Just north of the Flame Nebula lies a filamentary region exposed to the influence of the NGC\,2024 \Hii{} region. Further north, the round shape of the \Hii{} bubble becomes less visible ($\delta x;\delta y \approx -20\arcmin;-40\arcmin$). At the northern edge of our field lies another long filament, which we dub the Hummingbird filament. By eye, it is the second longest filament in the cloud after the Flame filament, and it stands out as an isolated structure, which makes it a perfect subject to study for example gravitational fragmentation.
Finally, to the north-east lies a blue-shifted turbulent region containing dense cores within a complex velocity structure. Here again, the filaments are not easy to identify by eye.

\subsection{Identifying the filaments}
\label{sec:extraction}

Visual inspection is insufficient to locate precisely and objectively and thus study filaments, in particular in the most entangled regions. Therefore, we implemented several algorithms to identify filamentary structures in a map. These algorithms are presented in detail in Appendix \ref{app:method}. We describe the concepts we use and we briefly summarise the algorithms that we apply to the \Nh{} map derived from the \COline{} integrated intensity.

%%%%%%%%
%In this paper, we restrict ourselves to study filamentary structures in a 2D map but the concepts and implementations can be generalized to a position-position-velocity cube.

\subsubsection{Morphological definition and extraction algorithms}

From an observational point of view, we can qualitatively define filaments as elongated, over-dense structures in the molecular interstellar medium (ISM). We thus expect to see these as bright structures with high aspect ratios. If we were looking at the altitude map of a mountainous region, where the altitude would correspond to the brightness of molecular emission, the filaments would correspond to the main mountain ranges of this region. The ridge lines of these mountain ranges would correspond in turn to the filamentary skeleton. These mountain ranges, and these ridge lines, can be simply defined in terms of topography, that is in terms of differential properties of the studied map.

The filaments are characterised by the properties of the Hessian matrix, the second derivative of the map \citep{schisano14}: the main directions of variation are identified and ridges appear as local maxima along one direction, while the perpendicular direction shows negligible, or at least smaller, variation (Table~\ref{tab:hess}). The first filament extraction procedure was meant to be as straightforward as possible, and simply applied a threshold on the eigenvalues of the Hessian matrix yielding the skeleton S1. The second procedure aimed at introducing several refinements: the data were rescaled with an arcsinh function prior to the computation of the Hessian matrix, then the eigenvalues were used to compute the local aspect ratio of the structures in the column density map, and the gradient was also used to refine the ridge detection. The whole analysis was performed in a multi-scale fashion, and yielded the skeleton S2. The details of both approaches are described in Appendix \ref{app:method}.

\subsubsection{Skeletons, bones, nodes, and filamentary network}
To avoid confusion between the different filamentary objects that are discussed in this paper, we define them as follows. The filament identification process yields a set of 1-pixel wide curves, which constitute a graph called the ``skeleton'' (or ``filamentary skeleton''). These lines have endpoints and intersections, which are called ``nodes''; the branches of the graph between two nodes are ``bones'' \citep[as in] []{panopoulou14}. The bones can be fleshed out by attributing some of the surrounding gas to each 1-pixel wide curve, thus yielding ``individual filaments''. Taken together, these individual filaments form a ``filamentary network'' (i.e. a fleshed-out skeleton). Some objects visible in the field and spontaneously referred to as filaments can be made of several individual filaments (e.g. the Flame filament); to avoid confusion these are therefore referred to as ``filamentary structures''. Finally, the term ``filamentary regions'' refers to the regions identified as bright and structured during the filament identification process, regardless of any attribution of the gas to a specific individual filament (binary masks shown in panel 5 of Fig. \ref{fig:filS1} and \ref{fig:filS2}).

\subsubsection{Skeleton analysis}

\begin{figure}
    \centering
    \includegraphics[width=\linewidth]{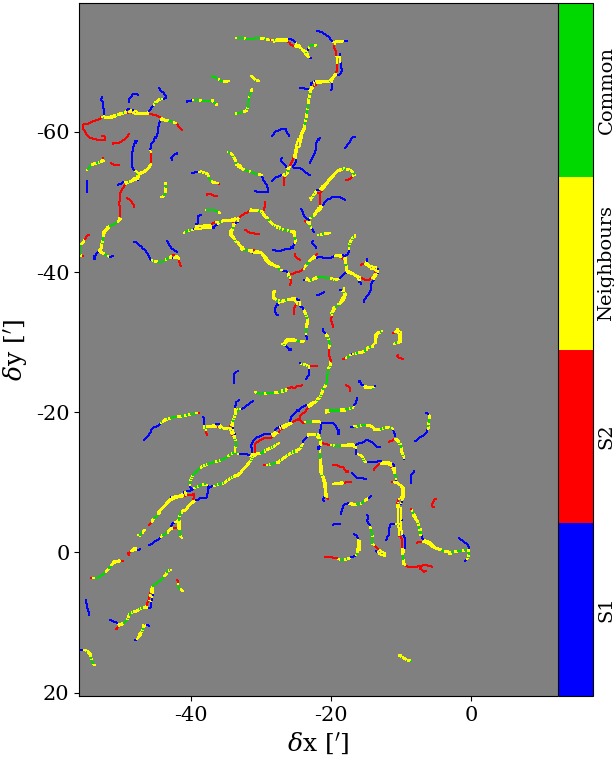}
    \caption{Comparison of the skeletons obtained with the two different methods. The S1 skeleton denotes the skeleton obtained by simple thresholding, and S2 is obtained by the adaptive method with ridge detection. The structures only identified by one method appear in red or blue, the structures common to both methods appear in green if they perfectly match, and in yellow in the cases where the morphological thinning led to small position offsets (see Appendix \ref{app:method}).}
    \label{fig:filcompare}
  \end{figure}

The two methods show disparities. Some structures can be identified by one method and not the other, or vice versa. To assess the similarities and differences quantitatively, we superimpose both skeletons (Fig.~\ref{fig:filcompare}). The different filtering approaches can lead to differences of the order of one or two pixels in the position of the identified structures. But as this does not have much impact on the further analysis of the filaments, such neighbouring points are considered as matching points, up to a distance of two pixels.  While there are some structures that are exclusive to one skeleton or the other, this criterion shows that a large portion (about 68\% of S1, 83\% of S2) of the skeletons is common to both methods.

Rather than choosing between the two possible skeletons, our approach is to keep both and study them and their statistical properties to assess the variability in the physical properties that can result from a variability in filament identification method. Therefore, four skeletons in total are compared throughout this work: the skeleton S1 obtained by simple thresholding; the skeleton S2 obtained by the adaptive method with ridge-detection; the ``robust skeleton'', which is made of the common or neighbouring points of S1 and S2 ,and thus corresponds to the intersection S1$\cap$S2; and the ``composite skeleton'', which is made of all the points of S1 and S2, and thus corresponds to the union S1$\cup$S2. Since the robust skeleton has, by construction, a thickness of two or three pixels because neighbouring points are taken into account, it needs to be thinned again (see Appendix \ref{app:extract}) to a single-pixel width. The first three skeletons are used for statistical comparisons, while the last is useful for displaying simple maps of the physical quantities.

%For that, a geometrical analysis allows us to distinguish between regular points and nodes (or vertices). The regular points have exactly two neighbours, while the nodes have fewer (if they are endpoints) or more (if they are intersections) than that. The individual filaments are therefore strings of points belonging to the skeleton and linking two nodes.

Once the skeletons are extracted from the observational data, we still need to identify the bones.  This sorting of the skeleton into bones and nodes allows us to analyse the local properties of the filamentary skeleton. It also enables us to ``clean'' (Appendix \ref{app:clean}) the skeleton by removing isolated nodes and short bones (under 0.22\pc\ long), and those that do not match the definition of what a filament should be: curvature (Appendix \ref{app:curv}) and contrast (Appendix \ref{app:contrast}) can show that some structures are not over-dense, narrow, or elongated enough to be regarded as filaments. With the exception of the Appendix, the figures and statistics presented in this paper are therefore obtained using the cleaned skeletons.

%The qualities and drawbacks of the two skeleton extraction methods are discussed in details in Appendix \ref{app:skel-quality}.

\subsection{Transverse profile fitting}
\label{sec:profile}
The Hessian identification of filaments provides us with their position angles. We can thus study the cross-sections (or transverse profiles) of the individual filaments, by plotting the variations of a physical quantity of interest perpendicular to the bone's local major axis -- in particular using the hydrogen column density map.

For the sake of simplicity and robustness (see \citet{arzoumanian11} and \citet{panopoulou14} for a discussion), the column density profiles of the individual filaments were considered to be Gaussian peaks superimposed on a linear baseline. Such a profile is therefore constrained by five parameters: the position $x_0$ of the Gaussian with respect to the reference pixel in the skeleton, the amplitude and width of the Gaussian, $A$ and $w$, and the slope and offset of the baseline, $\alpha$ and $\beta$ as follows:

\begin{equation}
P(x) = \alpha\cdot x + \beta + A\cdot\exp\left(\frac{-\left(x -x_0\right)^2}{2w^2} \right)
\label{eq:profile}
.\end{equation}
The range over which the profile is fitted, on either side of the skeleton ridge, can have a strong influence on the resulting model profile \citep{panopoulou14}. Therefore, we tried to use a spatial range as wide as possible, but we were limited by the density of the filamentary network: a very wide range can intersect several individual filaments, which leads to incorrect fitting results. As a compromise between these limitations, we set the fitting range to six pixels on each side of the ridge (or about 1.9') to limit this effect. We thus avoid getting a secondary bump in the outer parts of the profile, which would be due to a neighbouring individual filament. However, even with this fitting range, the same pixel in the column density map can be attributed to several profiles, either of the same individual filament (e.g. after a turn) or of neighbouring individual filaments.

%The chosen fitting function, given that it includes a linear baseline, can treat indifferently a \Nh{} column density map where the \discuss{2 \Av{}} contribution of the diffuse foreground and background has been subtracted or not, as discussed in Sect.~\ref{sec:c18o}.

We started by fitting the transverse profiles individually (i.e. for each pixel of the skeleton). However, this led to strong degeneracies between the various parameters when the baseline deviates from the assumed straight line, which required human supervision to be overcome. As wanted to avoid this in a semi-automated, statistical analysis of the filamentary network,  we set the individual filament width $w$ as a semi-fixed parameter. To determine its value, we fit the mean profile of each individual filament with all five free parameters, since this mean profile has a better \SNR{} and a more Gaussian shape than the profiles for individual pixels. The obtained value of $w$ is then kept for the transverse profile of each pixel in the individual filament. 

We have also taken into account the fact that the skeleton is not always perfectly aligned with the physical ridge lines of the filaments: as shown in Fig.~\ref{fig:filcompare}, the thinning of the skeletons can lead to positional uncertainties of about one or two pixels. For a given transverse profile, this means that the peak of the profile is off-centred, and that $x_0 \neq 0$. Such offsets can artificially broaden the mean profile. Therefore, a recursive approach is used: after a first fit of the mean individual filament profile which results in a given individual filament width, we fit each individual profile using that width as a fixed parameter. This individual fit yields in turn the position offset of the profile peaks, which allows us to re-align the profiles before recomputing an updated, centred mean profile. This centred mean profile is then fitted (with five free parameters) and yields a better (usually narrower) estimate of the average profile width in the individual filament. This filament is then used to perform a better fitting, with only $\alpha$, $\beta$, $A$ and $x_0$ as free parameters, of the individual profiles.

The fit results give us access directly to such quantities as the filament width (Sect.~\ref{sec:width}) and contrast (Appendix \ref{app:contrast}) and to quantities more closely linked to star formation, such as the mass and gravitational stability of the filaments (Sect.~\ref{sec:grav}).

\section{Physical properties of the filaments}
\label{sec:results}
\subsection{Profile and mass}
\subsubsection{Filament width}
\label{sec:width}
The characteristic width of filaments is a direct output of column density profile fitting, and it is the most commonly measured and discussed quantity for filaments in the ISM \citep[see e.g.][]{arzoumanian11, arzoumanian13, kainulainen16, panopoulou14, panopoulou17}. The filament width is measured on the global profile of each individual filament (as described above), using a non-weighted average; i.e. all profiles in the individual filament are normalised to the same amplitude. The obtained result is deconvolved from the synthetic Gaussian beam of the fitted IRAM 30\,m observations, which corresponds to 29\arcsec{} or about 60\,mpc. The final width is thus given by $w_\mathrm{deconv} = \sqrt{w_\mathrm{fit}^2-0.06^2}$\pc.

\begin{figure}
    \centering
    \includegraphics[width=\linewidth]{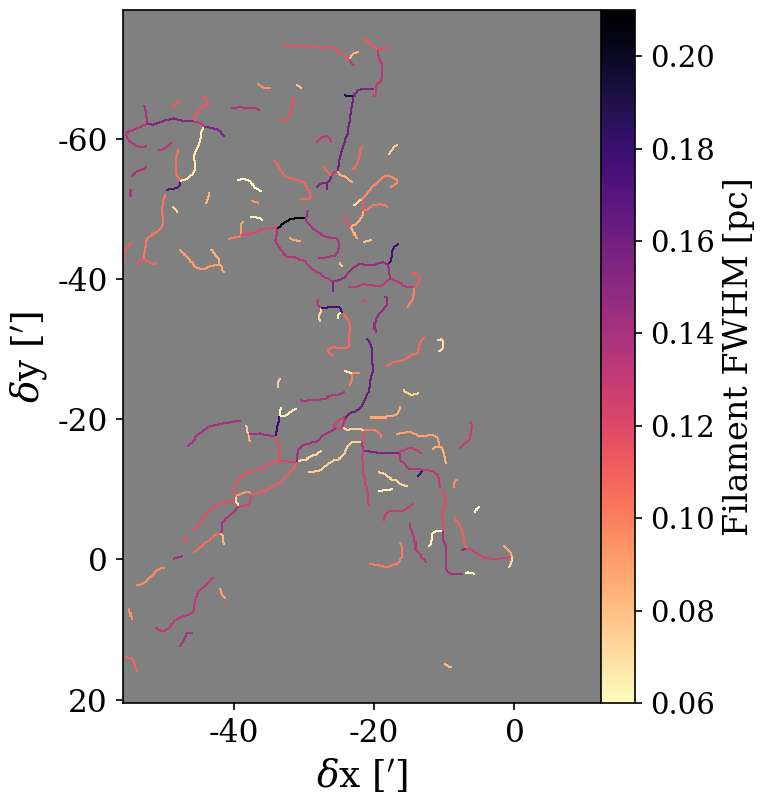}
    \includegraphics[width=0.9\linewidth]{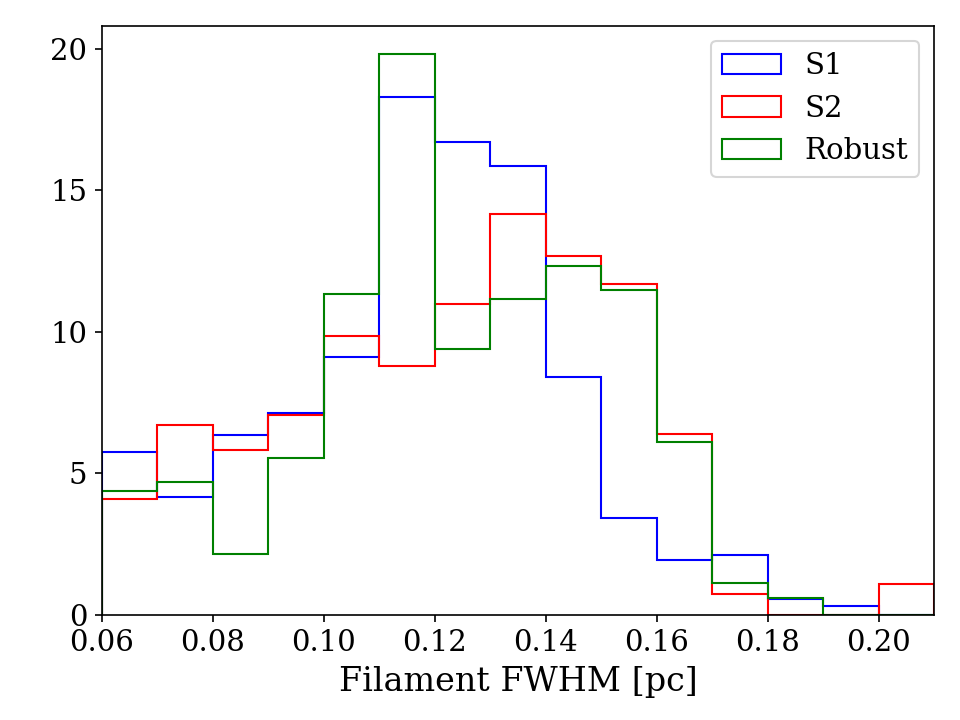}
    \caption{\emph{Top:} Map of the FWHM of the mean profiles of the individual filaments (combined skeleton). \emph{Bottom:} Histograms of the filament width for the S1, S2, and robust skeletons. }
    \label{fig:width}
  \end{figure}

The results are shown in Fig.~\ref{fig:width}. The typical widths for each skeleton, in terms of mean, median, or most probable value are listed in Table~\ref{tab:width}. Except for a few individual filaments, mostly short filaments for which the fit does not converge or yields oddly high values (which is the sign of filaments with low contrast, see Appendix \ref{app:contrast}), the spread of widths is rather small.

\begin{table}
    \centering
    \caption{Mean, median, and most probable FWHM of filaments for the three studied skeletons, resulting from mean profile fitting of the \CO{}-derived \Nh{} column density.}
    \begin{tabular}{ccccc}
      \hline
      \hline
      Skeleton  & Mean & Median & Most probable & Std dev\\
      \hline
      S1 & {0.11}\pc & {0.12}\pc & {0.13}\pc & 0.04\pc\\
      S2 & {0.11}\pc & {0.12}\pc & {0.13}\pc & 0.04\pc\\
      Robust & {0.11}\pc & {0.12}\pc & 0.11\pc & 0.04\pc\\
      \hline
    \end{tabular}
    \label{tab:width}
  \end{table}

\subsubsection{Linear density and gravitational stability}
\label{sec:grav}

The next physical quantity derived from the profile fitting is the linear density of the filaments. %The individual transverse profiles for each line of sight are fitted with the function defined in Eq. \ref{eq:profile} with a fixed width, which is the average width of the filament the line of sight belongs to.
The linear density $M_\mathrm{l}$ of a filament should take into account only the matter in the filament itself, not its foreground or background. This is why the linear baseline of Eq. \ref{eq:profile} is subtracted from the fitted profile. The linear density for a given line of sight is then simply the integral of the corresponding Gaussian transverse profile of surface density.

\begin{figure*}
    \centering
    \includegraphics[height=0.48\textheight]{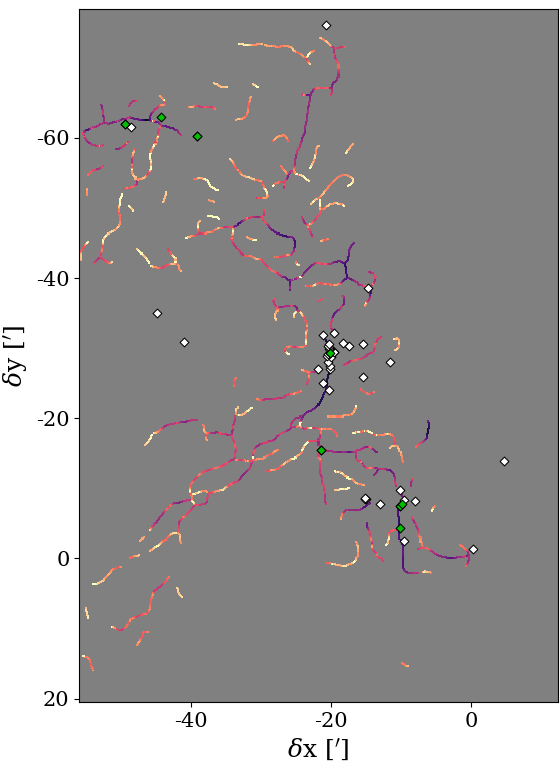} \hfill
    \includegraphics[height=0.48\textheight]{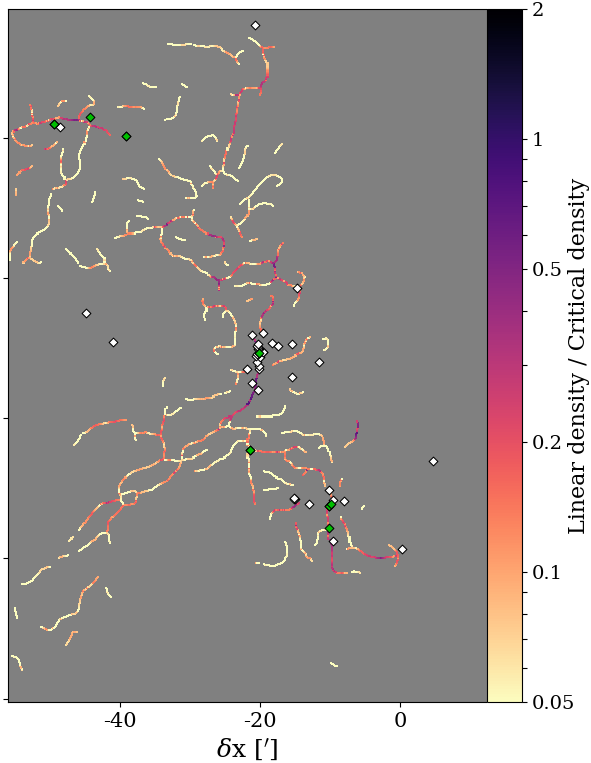}
    \caption{Gravitational instability criterion of the filaments (combined skeleton). This criterion is derived from the \CO{}-traced linear density estimation and either the thermal pressure ($\gamma$, \emph{left}) or the effective pressure, which also takes into account the \COline{} velocity dispersion ($\gamma_\mathrm{eff}$, \emph{right}). Overlaid are the positions of protostars from \citet{megeath16} in white, and class 0 YSOs from HOPS \citep{furlan16} in green.}
    \label{fig:stab}
  \end{figure*}

Knowing the linear density of the individual filaments and having a proxy for their kinetic  temperature $T_K$ thanks to the dust temperature map, we can also estimate the stability of Orion\,B's filaments against gravitational collapse. The criterion for balance between thermal pressure and gravity is given by $\gamma = M_\mathrm{l}/M_\mathrm{l}^\mathrm{crit}$, where $M_\mathrm{l}^\mathrm{crit}$ is determined by \citet{ostriker64} as
\begin{equation}
\label{eq:crit}
M_\mathrm{l}^\mathrm{crit} = \frac{2 k T_K}{\mu m_\mathrm{H} G} = \frac{2 c_\mathrm{s}^2}{G} \approx 16 \Msun \pc^{-1} \times \left(\frac{T_K}{10 \K{}}\right).
\end{equation}
The resulting gravitational instability criterion $\gamma$ of the filaments is presented in Fig.~\ref{fig:stab} (left).
 
However, with this approach, the critical linear density $M_\mathrm{l}^\mathrm{crit}$ is a lower limit because it only takes into account the thermal (kinetic) pressure. In order to account for support against gravity from both thermal and non-thermal motions of the gas, we need to compute an effective temperature $T_\mathrm{eff} = T_K + c_\mathrm{turb}^2\frac{\mu m_\mathrm{H}}{k}$, where $c_\mathrm{turb}$ is the observed non-thermal velocity dispersion \citep{arzoumanian13,peretto14,kainulainen16}.
We could have access to the turbulent velocity dispersion $c_\mathrm{turb} = \Delta v$ thanks to the velocity-weighted moments of the \COline{} spectra. However this would not take into account the fact that spectra can contain several velocity components. The spectral signature of the turbulence providing support in the form of an effective pressure is expected to be found in line broadening rather than in the multiplicity of spectral components, which rather trace the presence of physical substructures. Therefore, when computing $T_\mathrm{eff}$, the velocity dispersion $\Delta v$ that we use is rather the typical full width at half maximum (FWHM) of the Gaussian velocity components identified by the multi-Gaussian fit (Sect.~\ref{sec:profile}). The corresponding effective critical linear density $M_\mathrm{l,eff}^\mathrm{crit}$ is an upper limit this time. This is because of the implicit assumption that all the non-thermal spectral broadening arises from turbulent motions, which might not necessarily be the case; for example. accretion or gravitational collapse can also contribute to line broadening. From that, we derive the effective gravitational instability criterion $\gamma_\mathrm{eff}$ (Fig.~\ref{fig:stab}, right).

%\FigStabEff

%All the filaments in Orion\,B prove to be stable again radial gravitational collapse, with the stability criterion $\gamma$ barely reaching values of the order of 0.5.

Both the lower limit of the instability criterion, $\gamma_\mathrm{eff}$, and its upper limit, $\gamma$, show that the filaments in Orion\,B are mostly stable against gravitational collapse. This is further discussed in Sect.~\ref{sec:disc-stab}.

Since unstable filaments undergoing gravitational collapse are likely to lead to star formation \citep{arzoumanian11,hacar13}, we also compare the spatial distribution of the gravitational instability criteria $\gamma$ and $\gamma_\mathrm{eff}$ in the filamentary network with the positions of the youngest among the young stellar objects (YSOs) identified by \citet{megeath16} and by the \emph{Herschel} Orion Protostar Survey \citep[HOPS,][]{fischer13,furlan16}. We can see indeed on Fig.~\ref{fig:stab} that the positions of YSOs, in dense clusters (NGC\,2024), in looser groups (between NGC\,2023 and the Horsehead) or isolated, tend to correspond to local maxima of the instability criterion $\gamma$ or $\gamma_\mathrm{eff}$, i.e. regions where the filaments are closest to radially collapsing under the effect of self-gravity.

\subsection{Relative alignment of the magnetic field and filaments}
\begin{figure}
    \centering
    \includegraphics[width=\linewidth]{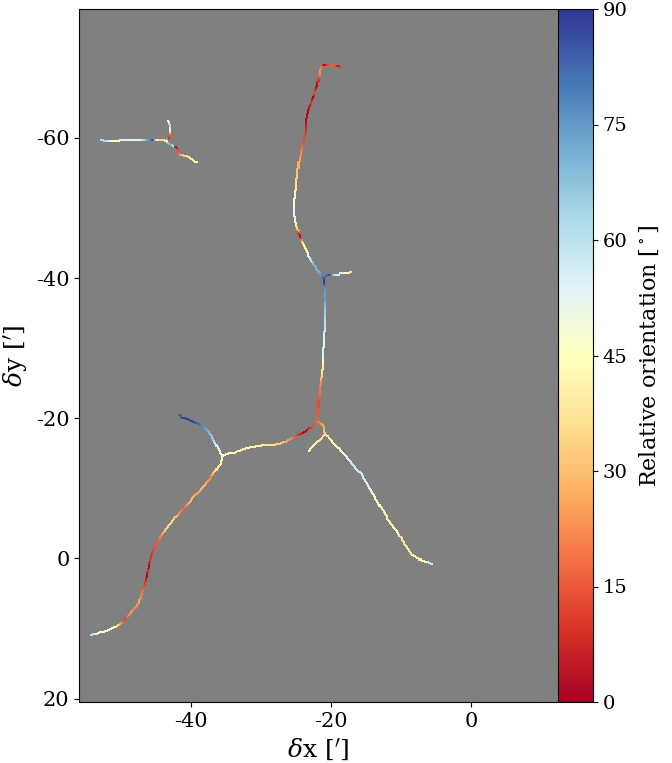}
    \includegraphics[width=\linewidth]{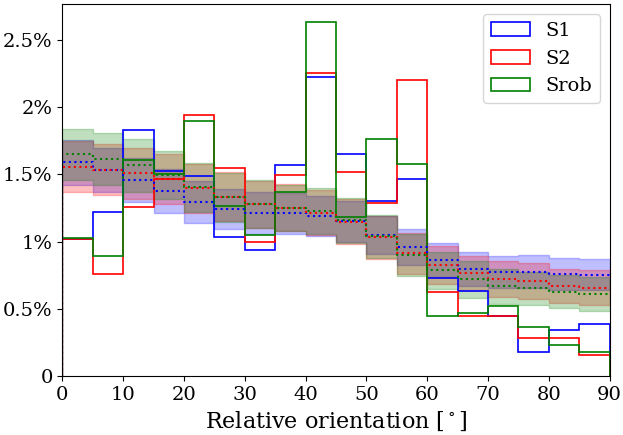}
    \caption{\emph{Top:} Map of the angle between the magnetic field and filaments detected in Orion\,B at a 5\arcmin\ resolution. The filaments are red when they are parallel to the magnetic field and blue when perpendicular to it. \emph{Bottom:} Distribution of the relative orientation of the filaments and the magnetic field for the S1, S2, and robust skeletons. The dotted lines (and the shaded areas) present the distribution (and the corresponding $\pm 1 \sigma$ uncertainties) that we would get if the two quantities were uncorrelated. The uncorrelated distribution is obtained by a Monte Carlo sampling of the magnetic field and filament position angles.}
    \label{fig:psiB}
  \end{figure}

The relative orientation of filaments with respect to the magnetic field is an important element of their dynamical evolution: sub-critical structures are expected to be parallel to the magnetic field, and super-critical structures perpendicular to it \citep{nagai98}. As we have access to an estimate of the magnetic field orientation thanks to \emph{Planck} data (Fig.~\ref{fig:data}), it is straightforward to compare this orientation with the position angle of bones, which is obtained as a by-product of the Hessian filament extraction as mentioned in Sect.~\ref{sec:profile}. Given that both the filament position angle and the orientation of the magnetic field are defined as only modulo 180\degr, their relative orientation is between 0\degr{} and 90\degr{} (Fig.~\ref{fig:psiB}). However, the major caveat is that the resolution of \emph{Planck} data does not match that of the IRAM 30\,m observations (5\arcmin{} and 31\arcsec{} respectively).% -- a problem not shared by most studies of correlations of the magnetic field and the massive structures of the ISM \citep[e.g.][]{PCi32}. 

We thus smoothed the IRAM 30\,m data to the resolution of the \emph{Planck} magnetic field data, i.e. 5\arcmin{}, and performed the filament extraction on that smoothed data, before comparing it to the magnetic field map. The distribution of this relative orientation is compared to what it would have been if the two orientations were uncorrelated. The uncorrelated distribution and corresponding error margins are obtained by a Monte Carlo sampling, which randomly associates one value from the observed filament position angle distribution with one value from the observed magnetic field orientation distribution (Fig.~\ref{fig:psiB}). %In the case of a non-weighted distribution of relative orientations, we can see few signatures of favoured geometries: there is a slight excess of points between 0\degr{} and 10\degr{}, and a small deficit between 15\degr{} and 25\degr{}. In the case of a distribution weighted by the linear densities, which highlights the behaviour of the densest and most visible filaments in the molecular cloud, we see a stronger deficit between 15\degr{} and 25\degr{}, and an excess between 35\degr{} and 55\degr{} approximately.

%\FigPsi

The distribution obtained for uncorrelated quantities is not flat, which is a result of the anisotropy of both the magnetic field (Fig.~\ref{fig:data}) and the large-scale structures of the gas. The actual distribution of relative orientation shows a modest peak around 20\degr{} and a major peak, which lies between 40\degr\  and 60\degr{}. The latter corresponds to the large filamentary structures with a position angle of roughly $\pm$45\degr{} (the north-eastern extension, part of the Flame filament, and mostly the NGC\,2023-Horsehead complex). However, as the resolution of \emph{Planck} data does not resolve the actual filamentary structures in the cloud, beam-averaging makes the distribution of relative orientation difficult to interpret. Polarimetry measurements at higher angular resolution are thus required to better understand how the magnetic field interacts with gas structures on the $\sim0.1\pc$ scale. Measurements of the dust continuum polarisation using the NIKA-2 \citep{catalano18,adam18} camera at the IRAM 30\,m telescope could provide a major improvement in that regard.

\subsection{Velocity field around filaments}

Compared to Herschel photometric images, the high spectral resolution of the molecular line data provides access to the motions of the gas in the filaments and their immediate surroundings.

\subsubsection{Line-of-sight velocity dispersion}
\begin{figure}
    \centering
    \includegraphics[width=\linewidth]{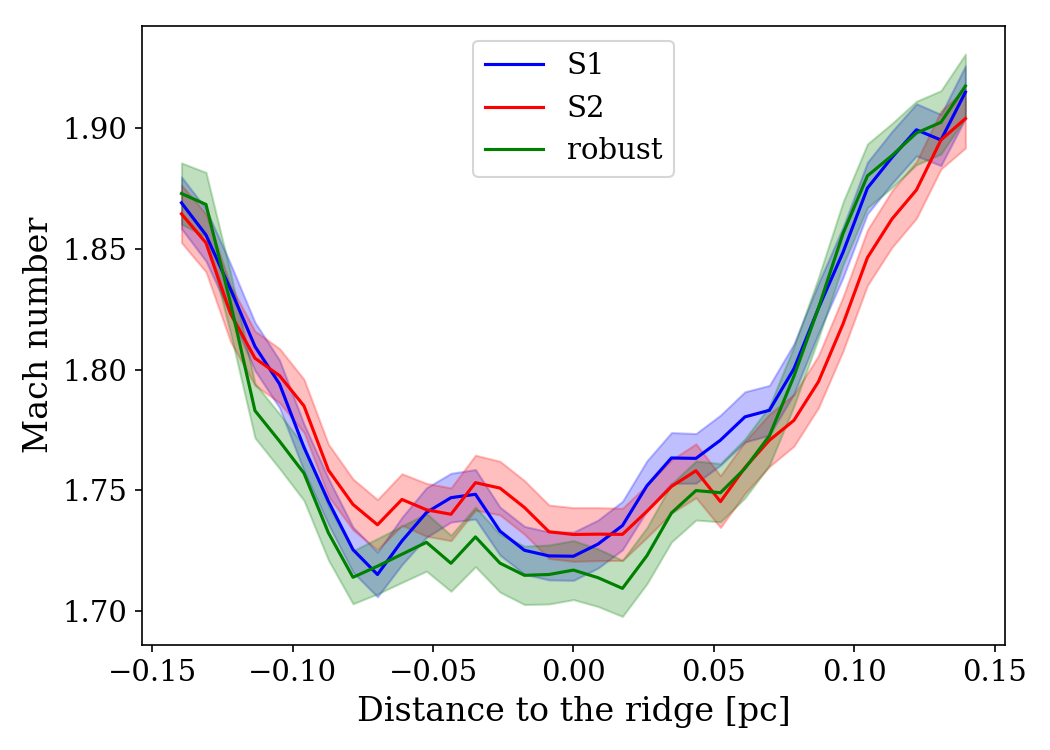} 

    \includegraphics[width=\linewidth]{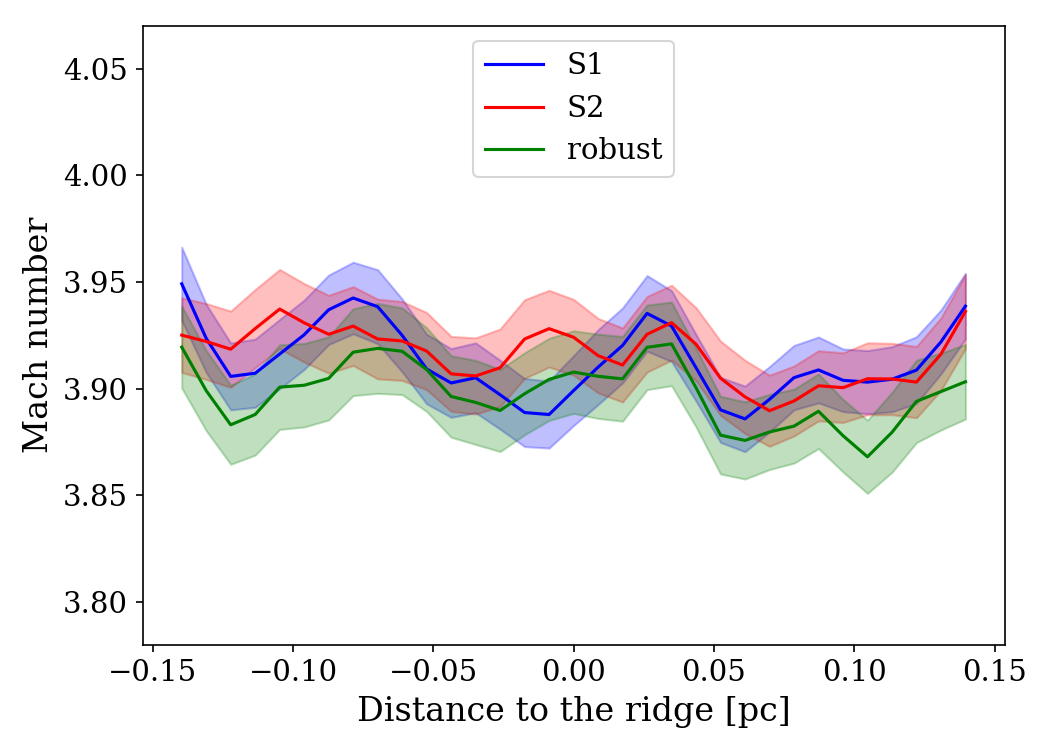}
    \caption{\emph{Top:} Average transverse profiles of the line-of-sight FWHM velocity dispersion of the filaments. These profiles are computed for the S1, S2, and robust skeletons via the modelled \COline{} linewidth; the shaded areas show the standard error for each profile. \emph{Bottom:} Same as above, but using the $^{13}$CO$\,(J=1-0)$ data cube for comparison.}
    \label{fig:dispersion}
  \end{figure}

As mentioned in Sect.~\ref{sec:grav}, we computed the line-of-sight FWHM velocity dispersion $\Delta v$ using the typical width of a Gaussian component identified by the multi-Gaussian fit. From this velocity dispersion we derived the Mach number, by comparing it with the sound speed $c_\mathrm{s} = \sqrt{\gamma k_\mathrm{B} T / m}$, where $T$ is the gas temperature (assumed equal to the effective dust temperature) and $m$ is the average molecular mass. The average transverse profiles of the Mach number in the filaments as probed by for \COline{} are shown in Fig.~\ref{fig:dispersion} (top). For comparison, we also plotted the transverse profiles of the Mach number as probed by $^{13}$CO$\,(J=1-0)$ in Fig.~\ref{fig:dispersion} (bottom). In that case, the gas temperature used to compute the sound speed was the composite temperature obtained using the effective dust temperature and the $^{12}$CO$(J=1-0)$ peak temperature. This takes into account the fact that $^{13}$CO is present in regions of lower density than \CO{}, where the gas and dust are not so well coupled. Under the assumption of LTE, the peak temperature of the (often strongly saturated) $^{12}$CO$(J=1-0)$ line can thus offer a better proxy for the kinetic temperature of the gas in the moderately dense envelope of the cloud. The details of the derivation of this composite temperature are found in \citet{orkisz17}. 

The $^{13}$CO$\,(J=1-0)$ profiles display significantly higher Mach numbers than \COline{}, but show no significant feature whatsoever. On the other hand, the \COline{} profiles show a pronounced decrease in Mach number towards the centre of the filaments. The width of this feature is similar to the measured filament width.

%\textbf{When weighting the velocity dispersion profiles by the linear density (which gives more importance to the areas with high linear densities, in particular in NGC\,2024, and thus gets closer to what would be obtained if limiting our study to dense, well-defined filaments), this behaviour is enhanced compared to the non-weighted average.}
%This behaviour is enhanced when weighting the profiles by the linear density (which gives more importance to the areas with high linear densities, in particular in NGC\,2024), compared to the non-weighted average.
%The boxcar-like increase of \COline{} velocity dispersions occurs over a width which corresponds very clearly to the width of the Gaussian column density profiles ($\sim 0.13\pc$).

\subsubsection{Centroid velocity gradient} 
While the velocity dispersion gives access to the kinematics along the line of sight, there is no direct way to observe velocity effects in the plane of the sky. As a proxy for such observations, we used the gradient of centroid velocity. Its amplitude measures the variations of the velocity field in the plane of the sky, in contrast to the linewidth, which probes the velocity dispersion along the line of sight.

Figure \ref{fig:cenprof} shows the average transverse profiles of the amplitude of the centroid velocity gradient observed in \COline{} (top) and in $^{13}$CO$\,(J=1-0)$ (bottom). The gradient amplitude is almost constant across the filaments in \COline{} with no particular visible trend. %But, given the amplitude of these variations (less than 0.5\kms/\pc{} on transverse scales of the order of 0.1\pc), velocity gradients in the plane of the sky appear to be about a factor 2 less intense than along the line of sight (0.1\kms\ increase in velocity dispersion on the same spatial scale).
The $^{13}$CO$\,(J=1-0)$ profiles, on the other hand, exhibit slightly higher amplitudes of the gradient, with a pronounced minimum towards the centre of the filaments, which brings the centroid velocity gradient amplitudes of  $^{13}$CO$\,(J=1-0)$ down to the average value observed for \COline{}.

\section{Discussion}
\label{sec:discussion}
\subsection{Width of the filaments}
\subsubsection{On the robustness of our width estimation}
\citet{panopoulou17} suggested that the narrow distribution of filament widths with a typical scale of 0.1\pc{} can be an artefact resulting from the analysis method, in particular the measure of the width of the mean profile of a filament rather than the widths of its individual profiles and the spatial range used for the profile fitting. The use of average filament profiles results in a narrower distribution of widths, but does not modify its peak significantly. In our case, the high spatial density of detected filaments and thus the number of intersections fragments the largest filamentary structures (such as the Flame filament) into shorter individual filaments, reducing the amount of averaging. The obtained distribution of filament widths, and in particular its dispersion, appears as an intermediate between the broad distributions obtained for individual profile widths and the very narrow distributions obtained for average filament profiles, as shown in \citet[][their Fig.~2]{panopoulou17}. The comparison with measurements on filaments in the Polaris Flare by \citet[]{panopoulou17} suggests that the absence of filaments with widths larger than 0.2\pc{} in Orion\,B can be a result of the fitting window, which is 1.9\arcmin{}, or 0.21\pc{} on each side of the filament ridge. However, the median and most likely widths in Fig.~\ref{fig:width} and Table~\ref{tab:width} are small enough to be confidently measured, but large enough not to be due to the telescope beam. We can therefore say that, even though the spread of widths of the filaments is probably underestimated in this work, the median and most probable value of the filament width, of the order of 0.12\pc, are reliable.

In addition to this first caveat, we should also stress that a Hessian detection filter behaves to some extent like a wavelet filter, bringing out structures matching the scale of the Gaussian derivative used for the calculation. However, the skeleton S2, thanks to its adaptive nature, can overcome this bias, since the data dictate the scales at which the filter has its strongest response (see Appendix \ref{app:extract}). For the skeleton S1, the use of a single smoothing scale could induce a stronger bias, but the robustness of the results was checked by studying skeletons obtained in the same way as S1, but with a halved or doubled smoothing scale. The distribution of the obtained filament widths however was slightly shifted towards smaller or larger scales, respectively, owing to the inability of a very narrow filter to pick up very broad structures and the excessive smoothing by a very wide filter that blurs out very small structures. The peak  of this skeleton remained unchanged, thus proving that the main detected structures do not strongly depend on the filter and that they correspond indeed to filaments with a mean width of the order of 0.12\pc{} (Table~\ref{tab:width}).

Finally, the average Mach number profiles derived from the \COline{} velocity dispersion show a feature of similar width (Fig.~\ref{fig:dispersion}). The width of the feature seen in the average profiles of the $^{13}$CO$\,(J=1-0)$ centroid velocity gradient norm is also of the order of 0.1 -- 0.2\pc. These measurements, completely independent from spatial distribution of the column density, argue in favour of the robustness of our estimate for the FWHM of filaments.

\subsubsection{Universal filament width?}
%The width, or diameter, of the interstellar filaments, is the first necessary measurement to obtain their linear density, and understand their evolution and their ability to form prestellar cores.

The distribution of mean filament widths that we obtain in Orion\,B, no matter which skeleton we consider, is very similar to that presented by \citet{arzoumanian11} for the IC\,5146, Polaris, and Aquila regions, where a 0.1\pc{} ``typical'' width with a 0.03\pc{} spread was reported, and to simulations results by for example \citet{federrath16}.

In contrast, our statistics of the filament widths are different from what is observed in the Taurus molecular cloud by \citet{panopoulou14}, based on $^{13}$CO$\,(J=1-0)$ observations \citep{narayanan08} that have a typical filament width of 0.4\pc{}. This discrepancy results at least partly from the chosen molecular tracer, which is less adapted for tracing the filamentary material, as discussed in Sect.~\ref{sec:obs-iram}. $^{13}$CO is susceptible to become optically thick and therefore to better trace the extended, power-law-like envelope of the filament rather than its tubular, central part \citep{arzoumanian11}, which can result in wider FWHM estimations. In the case of Orion\,B, the measurement of filament widths applied to the $^{13}$CO-derived \Nh\ column density yields filament widths of $0.18\pm0.04\pc$, which are wider than the $0.12\pm0.04\pc$ measured with \CO.

\begin{figure}
    \centering
    \includegraphics[width=\linewidth]{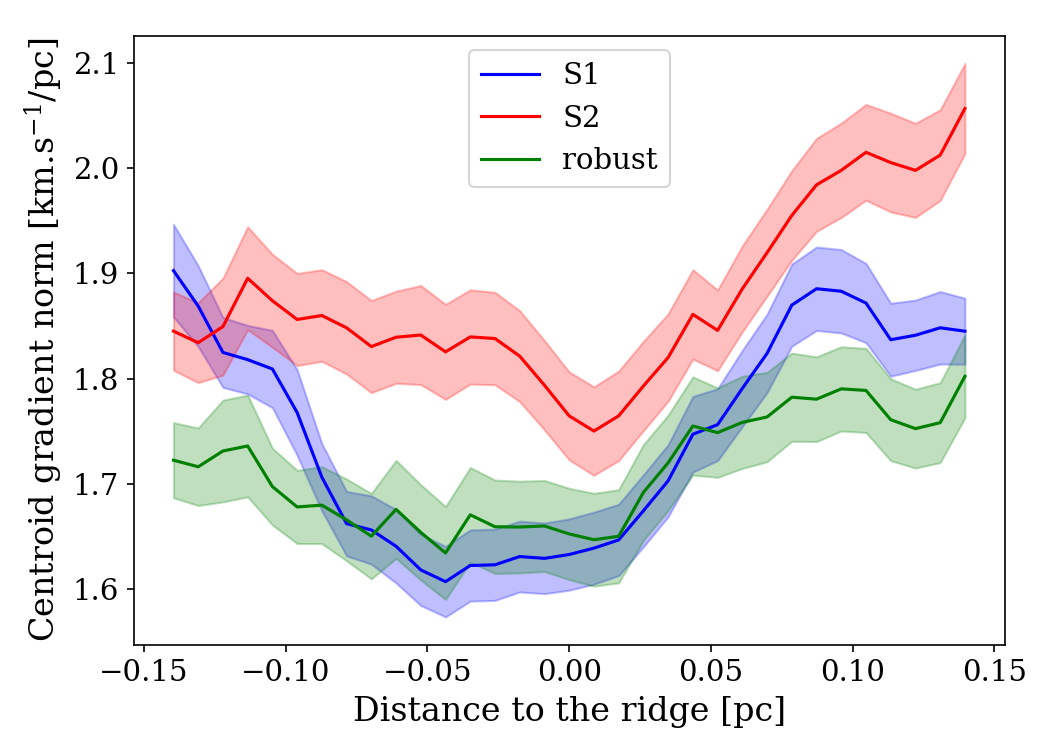}
    \includegraphics[width=\linewidth]{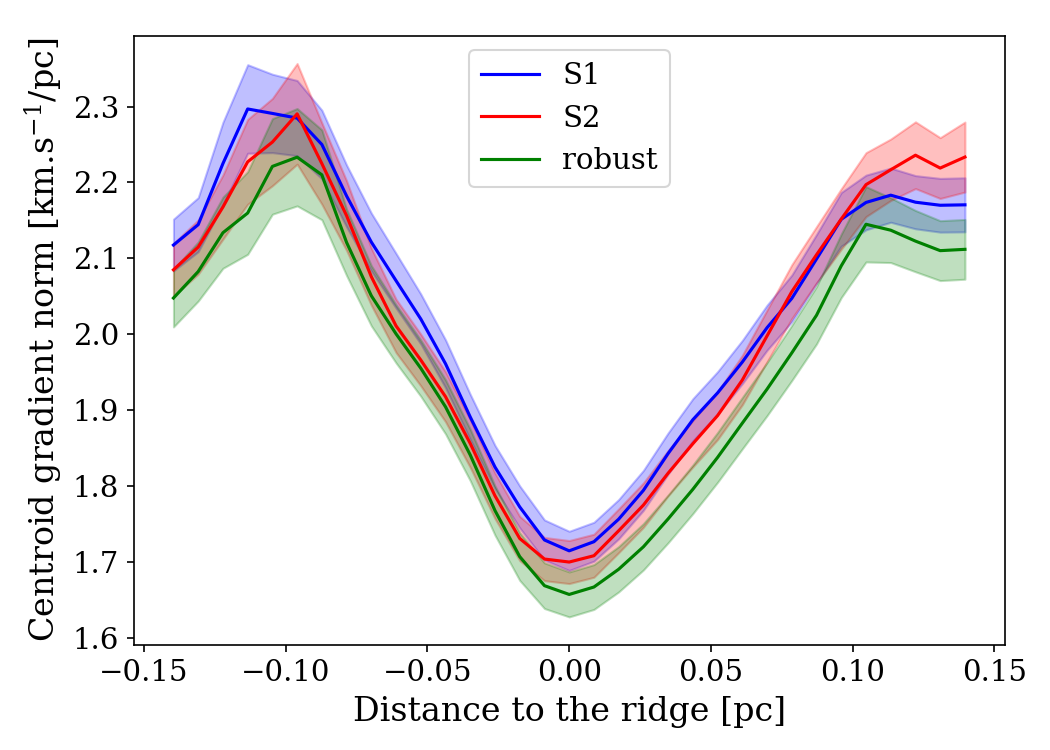}
    \caption{\emph{Top:} Average transverse profiles of the amplitude of the centroid velocity gradient of the filaments. These profiles are computed for the S1, S2, and robust skeletons via the \COline{} data cube; the shaded areas show the standard error for each profile. \emph{Bottom:} Same as above, but using the $^{13}$CO$\,(J=1-0)$ data cube for comparison.}
    \label{fig:cenprof}
  \end{figure}

\subsection{Gravitational stability and star formation}
\label{sec:disc-stab}
\subsubsection{Low linear and volume densities}

\begin{figure}
    \centering
    \includegraphics[width=\linewidth]{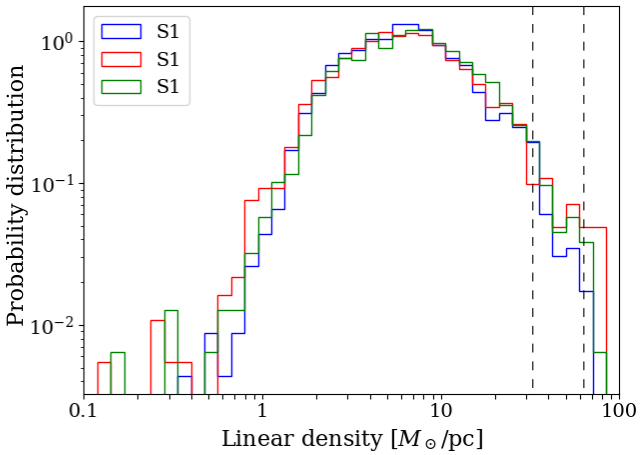}
    \includegraphics[width=\linewidth]{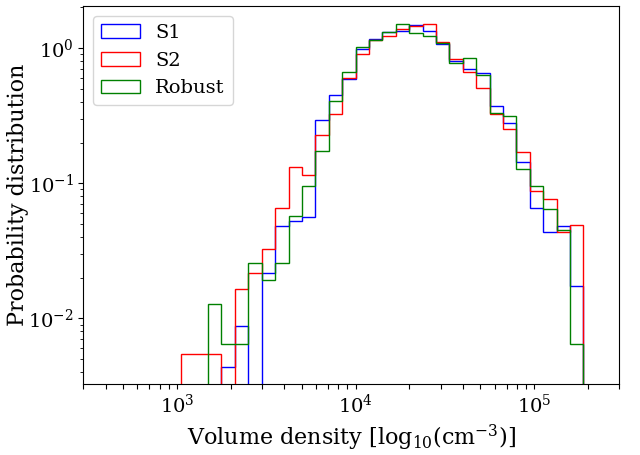}
    \caption{\emph{Top:} Distribution of the linear density of the filaments. The vertical dashed lines show the most probably critical linear densities corresponding to $M_\mathrm{l}^\mathrm{crit}$ and $M_\mathrm{l,eff}^\mathrm{crit}$. \emph{Bottom:} Distribution of the volume density of the filaments, derived from the linear density and the FWHM of the transverse profiles by assuming a uniform cylindrical geometry. The typical critical volume densities lie beyond the plot, at $\gtrsim 3\e{5}\pccm$.}
    \label{fig:linvoldens}
  \end{figure}

The linear densities of the filaments that we detected range from a few $\Msun/\pc$ to about $100 \Msun/\pc$, with a median linear density of $\sim 5 \Msun/\pc$ for all skeletons (Fig.~\ref{fig:linvoldens}, top). This apparently log-normal distribution matches the usual linear densities of interstellar filaments rather well, as observed for example in IC 5146 \citep{arzoumanian11}, Taurus \citep{panopoulou14}, or Musca \citep{kainulainen16}. This distribution is of course lower than what is observed for high linear-density filaments such as the Integral-Shaped Filament \citep{stutz15,kainulainen17}, for which the linear density is of the order of several $10^2\Msun/\pc$. More precisely, while the upper end of our distribution reaches the typical order of magnitude for linear densities (in the range of a few tens or hundreds of $\Msun/\pc$), a large fraction of the filamentary network have rather low linear densities.

This is also visible when looking at the typical volume densities of the filaments, which are estimated by assigning the linear density to a uniform cylinder, the diameter of which would be the FWHM of the individual filament profile. These volume densities range from $10^4$ to $10^5$\pccm, again with a distribution close to a log-normal one, with a median value of $\sim 2\times 10^4$\pccm (Fig.~\ref{fig:linvoldens}, bottom). This is consistent with the upper end of the volume density distribution in the whole western edge of the Orion\,B molecular cloud, as presented in \citet{bron18}. These typical densities are however lower by an order of magnitude than those measured by \citet{teixeira16} (and references therein) or \citet{kirk15}, based on observations of filaments or hydrodynamical simulations, respectively.

These distributions of linear densities and volume densities can be affected by completeness effects. The chosen molecular tracer, \CO, has broad sensitivity range \citep{pety17}, but it still has its limits,; this implies that faint structures lying below $~\sim 2\e{21}\pscm$ can be missed, while the densest filaments can have their density underestimated due to CO freeze-out or line saturation (Sect. \ref{sec:c18o}). In addition, the filament identification process removes a number of low-contrast individual filaments (Appendix \ref{app:contrast}) which would have fallen into the lower end of the density distributions. In total, completeness effects might affect low-density filaments more than high-density filaments.

We therefore have a set of filaments which contains a few objects matching the usual linear or volume densities found in the literature, but with an excess of low-density elements. There are several possibilities to explain this effect. First, \citet{andre10} and \citet{arzoumanian11} noted that fitting filament profiles with a Gaussian rather than a Plummer profile can lead to an underestimation of their density by about 20\%. However, this does not increase the densities we measured by an order of magnitude, and many studies cited above also used Gaussians to fit the transverse profiles. A more important factor is the detection scheme used in this work. Most studies focus on a small number of well-identified and carefully selected filaments, for example by setting high persistence levels when detecting filaments with DisPerSE \citep{sousbie11}. In contrast, we use a lax definition of the filaments, resulting in a number of rather faint, but still contrasting and elongated objects to be part of the analysed skeletons. In certain cases, this is desirable, for example in the case of the south-eastern extension of the Flame filament, which clearly divides into many substructures that we do not want to miss. In other cases, structures that would usually not classify as filaments are retained, such as lower density striations; i.e. strands of more diffuse gas that is accreting onto a main filament.

\subsubsection{No signs of gravitational collapse}

The filaments in Orion\,B are striking owing to their exceptional stability against gravitational collapse, even when looking at the higher limit of their instability criterion (Fig.~\ref{fig:stab}, left). While the denser filaments in our skeletons match rather well the sample in \citet{arzoumanian13}, this exceptional stability of the filaments can be explained by several factors.

First, the western edge of Orion\,B is a warm environment, heated by the large amount of FUV radiation coming both from outside ($\sigma$\,Ori) and inside (NGC\,2024) the cloud. These high temperatures (Fig.~\ref{fig:data}, right) lead to high thermal pressures and therefore high critical masses (Eq. \ref{eq:crit}). \citet{arzoumanian13} simply assumed a constant temperature of 10\,K in the filaments, \citet{teixeira16} a constant temperature of 15\,K, while the dust temperature in Orion\,B rarely drops below 20\,K. These warm temperatures could either be a layering effect or an actual specificity of this region of Orion\,B; in the layering effect, the filaments are actually
cold but the dust temperature is dominated by the warmer surrounding medium. In any case, an overestimated or genuinely higher temperature of the gas leads to higher critical linear densities.

Second, turbulence also plays a role in stabilising the filaments against gravity. \citet{kainulainen16} showed that taking turbulence into account brings the Musca filament from a super-critical to trans-critical state. In the case of Orion\,B, this corresponds to the dramatic difference between the left and right panels of Fig.~\ref{fig:stab} -- although, as mentioned, $\gamma_\mathrm{eff}$ is a lower limit, as some of the velocity dispersion might come, for example from infall / collapse \citep{arzoumanian13}, and not from rotation or turbulent motions that would support the filaments against gravity.
Table~\ref{tab:stab} summarises by how much the fraction of the filamentary network prone to gravitational instability would have increased if we had assumed no turbulent support or a lower gas temperature. It shows how important velocity-resolved observations of molecular lines are when trying to determine the stability of filaments.

\begin{table}
    \centering
    \caption{Fraction of super-critical ($\gamma > 1$) or at least trans-critical ($\gamma > 0.5$) filaments depending on the assumptions on their internal pressure computed for the S1, S2, and robust skeletons.}
    
    \begin{tabular}{cccc}
      \hline
      \hline
        & $T = 10\K$ & $T=T_K$ & $T = T_K + T_\mathrm{turb}$ \\
      \hline
      \multicolumn{1}{c}{} & \multicolumn{3}{c}{S1 skeleton} \\
      \hline
      $\gamma > 1$ & 9.9\% & 0.4\% & 0.1\% \\
      $\gamma > 0.5$ & 33.2\% & 8.0\% & 0.4\% \\
      \hline
      \multicolumn{1}{c}{} & \multicolumn{3}{c}{S2 skeleton} \\
      \hline
      $\gamma > 1$ & 9.3\% & 1.2\% & 0.2\% \\
      $\gamma > 0.5$ & 27.2\% & 7.4\% & 0.8\% \\
      \hline
      \multicolumn{1}{c}{} & \multicolumn{3}{c}{Robust skeleton} \\
      \hline
      $\gamma > 1$ & 9.0\% & 0.6\% & 0.1\% \\
      $\gamma > 0.5$ & 25.4\% & 7.6\% & 0.4\% \\
      \hline
    \end{tabular}
    \label{tab:stab}
  \end{table}

%\discuss{It is interesting to note that the width of the filaments and their gravitational stability seem to be only weakly correlated (Fig.~\ref{fig:widthstab}).} \jocomment{remove ?}

%\FigWidthStab

This overall lack of gravitationally unstable filaments in Orion\,B correlates well with its known low SFE \citep{lada92,carpenter00,federrath13,megeath16,orkisz17}. Moreover, the NGC\,2023 and NGC\,2024 star-forming regions are among the few regions containing super-critical or trans-critical filaments. This is consistent with the fact that these regions also show the most compressive motions, as measured by \citet{orkisz17}. 

\subsubsection{Star formation efficiency}
Measurement of the filament masses also enables us to check what fraction of the mass of the molecular cloud is contained in the filamentary network, and, in particular, in the gravitationally unstable filaments, as this last fraction directly relates to the SFE of the cloud. Using the $^{12}$CO$\,(J=1-0)$ line and following \citet{solomon87} and \citet{bolatto13} in the same way as in \citet{pety17}, we obtain a total virial mass of the cloud (for the considered field of view and accounting for the CO-empty IC\,434 PDR) comprised between 8400 \Msun and $13900\Msun.$ For simplicity, we use the average of these values, at 11100\Msun. This yields a fraction of mass in the filamentary network of about $4.3\pm1.1$\% ($474\pm28\Msun$) in the case of the S1 skeleton, $3.6\pm0.9$\% ($405\pm22\Msun$) in the case of S2, and $3.2\pm0.8$\% ($357\pm21\Msun$) for the robust skeleton. It is significantly less than the fraction of mass derived for the ``environment of filaments'' in \citet{pety17} (about 40\%), which is mostly explained by the sparse character of the filamentary network, compared to an \Av\ extinction mask, and also less than the total mass traced by \CO\ (about 1200 \Msun, or $11\pm3$\% of the mass of the cloud). Thus, only about one-third of the \CO{}-traced mass is found in filaments.

The fraction of mass in gravitationally super-critical or trans-critical filaments depends on the definition of the instability criterion. When using $\gamma$, we have $266\pm17\Msun$, or $2.4\pm0.6$\% of the mass of the cloud in trans-critical filaments, and $111\pm7\Msun$, or $1.0\pm0.3$\% of the mass in super-critical filaments. When using $\gamma_\mathrm{eff}$, the fraction of the mass of the cloud in trans-critical filaments is $0.9\pm0.2$\% ($95\pm6\Msun$), with only $0.1$\% of the mass ($21\pm1\Msun$) in super-critical filaments. To first order, the fraction of the mass of a molecular cloud contained in gravitationally unstable structures can be directly related to the SFE of that cloud, as we can roughly assume that this mass is going to collapse into cores and into stars. Among the fractions mentioned above, the fraction of the mass of the cloud in the trans-critical filaments using the gravitational instability criterion $\gamma$ can be considered as the upper limit for the SFE. Such a value of $2.4\pm0.6$\% is consistent with previous measurements of the SFE by \citet{lada92}, \citet{carpenter00}, \citet{federrath13}, or \citet{megeath16}, which range from 0.4\% to 3\%.

\citet{lada10} proposed that the star formation rate of molecular clouds is proportional to the mass above a threshold of 0.8\,\Ak , which corresponds approximately to 6 magnitudes of \Av, or $1.1\e{22}\pscm$, while \citet{andre14} find prestellar cores only in regions with column densities higher than $1.4\e{22}\pscm$ (i.e. $\Av\approx8\,$mag). These thresholds are higher than the detection limit of \COline{} (about 3 magnitudes of \Av, as mentioned in Sect.~\ref{sec:obs-iram}, which corresponds to $\approx 0.4$ magnitudes of \Ak), and exclude part of the filamentary regions (Fig.~\ref{fig:data} left, Fig.~\ref{fig:nhco}). They delimit a star-forming mass of the cloud of $\sim 540\Msun$ ; this mass is less than the total of the \CO-traced mass, but much more than what is contained in trans-critical or super-critical filaments. The typical volume density threshold that we obtain for gravitationally unstable filaments, using a typical value of the critical linear density of 30 -- 60$\Msun/\!\pc$ (Fig.~\ref{fig:linvoldens}, Eq. \ref{eq:crit}) and assuming a filament diameter of 0.12\pc, is about $5\e{5}\pccm$. It corresponds to a column density of $\sim 4\e{22}\pscm$, which is significantly higher than the threshold column densities proposed by \citet{lada10} and \citet{andre14}. The $\sim 1.1\e{22}\pscm$ threshold can thus be seen as a general absolute minimum under which the gas contained in molecular clouds does not contribute to star formation, but the actual column density threshold over which star formation is likely to occur in a given cloud can depend on intrinsic properties of the cloud, such as its age or kinematics. In the case of Orion\,B, in which strong solenoidal motions lower the SFE \citep{orkisz17}, this actual threshold is particularly high.

\subsubsection{One clear example of longitudinal fragmentation}

When trying to identify mechanisms of star formation in the filamentary network, it is difficult to conclude in favour of radial collapse, or longitudinal collapse leading to star formation via accretion onto hubs \citep{peretto14} based on the relative positions of the YSOs and the filamentary skeleton (Fig.~\ref{fig:stab}). However, another widely observed phenomenon leading to the formation of prestellar cores is longitudinal fragmentation \citep{hacar13,teixeira16,kainulainen16,hacar18}. 
Filaments that are above the gravitational instability limit are expected to collapse radially. But when they come close to their gravitational stability limit ($\gamma \lesssim 1$), filaments are mostly susceptible to fragment longitudinally on scales close to their Jeans length \citep{takahashi13,teixeira16} or to four times their FWHM \citep{inutsuka92}. 

\begin{figure}
    \centering
    \includegraphics[width=0.9\linewidth]{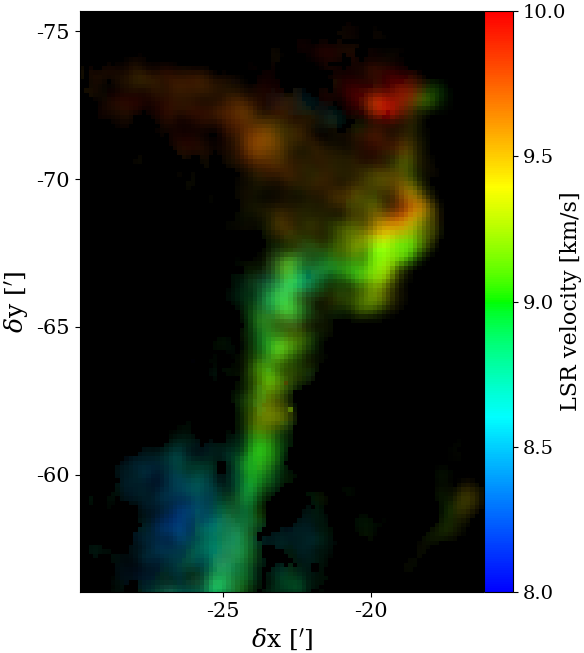}
    \includegraphics[width=\linewidth]{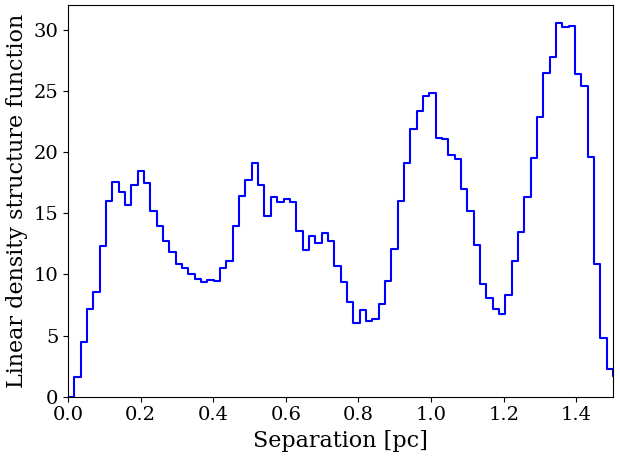}
    \caption{\emph{Top:} Same as Fig.~\ref{fig:rgb}, but with the field and velocity range zoomed in on the Hummingbird filament, showing beads in the rectilinear tail of the Hummingbird. \emph{Bottom:} Two-point structure function of the linear density in the Hummingbird filament (following the robust skeleton), in $\left(M_{\sun}/pc\right)^2$, showinging a periodic pattern which suggests longitudinal fragmentation with a period of 0.4\pc.}
    \label{fig:long}
  \end{figure}

Signs of such fragmentation are visible in at least one individual filament in our field -- the Hummingbird filament, which, interestingly, does not harbour any known YSO. Oscillations of the linear density are visible along this filament, forming evenly spaced ``beads". Following the recommendations of \citet{schulz81}, we computed the two-point auto-structure function of the linear density along the curvilinear abscissa of the filament's bone. It is defined as follows:
\begin{equation}
F(u) = \left\langle \left(\lambda(x) - \lambda(x+u) \right)^2 \right\rangle_x,
\end{equation}
where $\lambda$ is the linear density, $x$ a position along the bone, and $u$ the separation between the considered points. The resulting function is plotted in Fig.~\ref{fig:long}.

The oscillating pattern highlights the presence of evenly spaced structures with sizes of the order of 0.4\pc. Given that we have a FWHM of 0.11\pc{} for that individual filament, such a fragmentation length matches well the analytical prediction of \citet{inutsuka92} for collapsing isothermal filaments, and the simulation results of \citet{clarke16} for accreting filaments, since both predict a fragmentation with separations of about four times the diameter.

Following \citet{spitzer98}, \citet{takahashi13} and \citet{teixeira16}, we computed the Jeans length of the filament, $l_\mathrm{Jeans} = \sqrt{(\pi\,c_\mathrm{eff}^2)/(G\,n_0)}$, where $c_\mathrm{eff}$ is the effective sound speed, corresponding to the effective temperature , and $\rho_0$ is the volume density. The mean volume density $n_0 = 1.5\e{4}\pccm$ of the individual filament, derived for the case of a uniform cylinder, combined with an effective temperature $T_\mathrm{eff} = 43\K$ (while $T_K = 18\K$) yields a Jeans length of 0.54\pc, which is slightly higher than the observed characteristic scale in Fig.~\ref{fig:long}. We obtain a Jeans length of
0.38\pc \ if we replace this average density by an estimate of the maximum density that is obtained by comparing the average value of a Gaussian over its FWHM to its peak value and correcting for the transition from two to three dimensions as follows:  $\rho_{0,\mathrm{max}} = \left\langle\exp(-x^2/2) \right\rangle_\mathrm{FWHM} ^{-3/2}\cdot\rho_0 \approx 1.4\cdot\rho_0$. The observed fragmentation length is therefore close to the scales expected from both a cylindrical or a spherical instability.

In summary, filaments in Orion\,B show no signs of radial collapse, no clear evidence (at this stage of the analysis) of longitudinal collapse onto hubs, but at least one good example of longitudinal fragmentation.

\subsection{Kinematics}

The $^{13}$CO$\,(J=1-0)$ line traces the moderately dense gas that forms the bulk of molecular clouds and surrounds the filamentary network. Towards filament ridges, a substantial fraction of the $^{13}$CO$\,(J=1-0)$ emission can start to originate from the filaments, traced by the \COline{} emission. The kinematics of $^{13}$CO could thus trace the transition between the turbulent environment and the quiescent inner part of filaments \citep[e.g.][]{hatchell05,federrath16}. Towards the centre of the filaments, the norm of the centroid velocity gradient decreases, reaching at its minimum the same value as for \COline{}.
We would expect the Mach number to decrease as well, but it shows flat profiles instead. This could be explained by the fact that the optical depth of $^{13}$CO$\,(J=1-0)$ increases towards the centre of the filaments, resulting in a non-negligible opacity broadening which compensates for the decrease in linewidth due to a lower velocity dispersion. Indeed, \citet{orkisz17} have shown that the $^{13}$CO$\,(J=1-0)$ opacity broadening is negligible except in the dense regions in which we are interested. For the column densities of $^{13}$CO typical of the densest filaments ($N_{^{13}\mathrm{CO}} \sim 1\e{17}\pscm$), the opacity can reach $\tau_{^{13}\mathrm{CO}} \gtrsim 3$ and lead to a line broadening by a factor $\sim 1.5$; this is consistent with the absence of variation of the FWHM across the filaments because the decrease in velocity dispersion is
compensated by an increase in opacity broadening. The rather low Mach number traced by $^{13}$CO$\,(J=1-0)$ compared to the average of $\sim6$ given in  \citet{orkisz17} is because we only take into account one of the multiple spectral components along the line of sight, which can significantly reduce the measured velocity dispersion \citet{orkisz17}.

The \COline{} results contrast with $^{13}$CO$\,(J=1-0)$. The velocity dispersion \emph{along the line of sight} is supersonic, however the Mach number decreases significantly inside the filaments, down to almost transonic values, as it is usually observed or predicted \citep[e.g.][]{arzoumanian13,federrath16}. This can be a sign of the dissipation of turbulence \citep{hennebelle13}. The norm of the centroid velocity gradient possibly shows a similar, but far less pronounced decrease towards the centre of the filaments. We also notice that with filament widths of the order of 0.1\pc{} and centroid velocity gradients of $<2\kms{}/\pc{}$, the centroid velocity variations \emph{in the plane of the sky} are subsonic, given typical sound speeds of the order of 0.3--0.7\kms{}. This is reminiscent of the results of \cite{smith16}, who show that filaments are structures which move coherently on large scales (and thus have near-constant centroid velocities), regardless of the small-scale turbulence.

%One possible explanation of this discrepancy between the line-of-sight and the plane-of-the-sky behaviours might be the presence of unresolved \emph{fibres} \citep{hacar13} in most of the Orion\,B filaments. Such fibres, with sub-beam FWHM \citep{hacar18}, would have \emph{on average} the centroid velocity of the filament, and would have transonic velocity dispersion. As a result, the variation of centroid velocity of the filament in the plane of the sky would be transonic or subsonic, but the superposition in a single beam of several fibres with slightly different individual centroid velocities would result in a broadening of the \COline{} line. Since these fibres are expected to be denser and thinner than a 0.1\pc{}-wide filament, one can expect to have a sharper edge of this velocity dispersion effect, and therefore an effective tool to measure the actual width of the filament.

In the context of the velocity field of filaments, one element often discussed in the presence of fibres \citep{hacar13,panopoulou14,hacar18}.
We would expect from the possible presence of unresolved fibres that there would be multiple spectral components detected in many lines of sight in the filaments. However, it appears to be rarely the case. Only the part of the Flame filament closest to the NGC\,2024 star-forming region displays a long portion that consistently has two or more identified spectral components. %It must be stressed that the \COline{} velocity dispersion on Fig.~\ref{fig:dispersion} is based on the ``typical'' peak width deduced from the model cube, not on the second spectral moment (which would have shown an even stronger increase of the velocity dispersion, especially in the case of the linear-density weighted profiles which are dominated by NGC\,2024).
This calls into question the presence of fibres since it would mean that they are not only unresolved spatially, but also spectrally, since in most cases their presence would merely broaden the \COline{} line instead of showing more spectral peaks. Their presence could then possibly explain partly the supersonic Mach numbers even at the centre of the filaments.

A further understanding of the dynamics of these filaments using the ORION-B dataset would require performing a three-dimensional identification of the structures, which would improve the understanding of the spectral multiplicity of the filaments and provide a better view of their crossings and hubs. Further research on the velocity not only across but also along the filaments, studying their position-velocity diagrams in a fashion similar to for example \citet{peretto14}, would also benefit our understanding of these filaments. For example, the observed longitudinal fragmentation in the Hummingbird filament appears to be associated with a specific velocity pattern that will be explored in detail in a future paper.

\subsection{Statistical influence of the filament extraction method}

The study presented in this work shows that the difference between the statistical results of the different skeletons are modest, if not negligible. This could be expected from the significant overlap of the skeletons, which is in turn related to the consistent manner in which the tuning of the detection parameters was done, be it for the initial extraction (Appendix \ref{app:extract}) or during the cleaning process (Appendix \ref{app:clean}), for both the S1 and S2 skeletons.

When comparing Fig.~\ref{fig:filcompare} and \ref{fig:stab}, we can see that most of the individual filaments that are exclusive to either S1 or S2 are particularly stable, i.e. they are among the least dense structures in the field of view. The only significant structures that are not in common are the Horsehead PDR and a small portion of the Flame filament in NGC\,2024. This last point can explain the difference in percentage of mass in super-critical or trans-critical filaments observed between S1, S2, and the robust skeleton, as the missing fragment of NGC\,2024 in S1 contains a significant amount of mass above the gravitational instability threshold. The similarity of the results obtained with the robust skeleton compared to those with the S1 and S2 skeletons is also important, because it shows that the structures not detected by either S1 or S2 do not play a major role in the statistical properties of the filamentary network.

Provided that the definition of a filament is consistent with what was used in this paper (in particular in terms of density or length requirement) it is safe to assume that filament detection schemes do not play a significant role in the difference between different studies of interstellar filaments.

\section{Conclusions}
\label{sec:conclusion}

In this paper, we have used velocity-resolved line maps obtained through IRAM 30\,m observations in the context of the ORION-B project to trace the dense, filamentary matter in the Orion\,B molecular cloud with the \COline\ line. Using two different extraction methods, we identified the network of filaments in the cloud and used both extracted skeletons as well as their intersection to compare the statistical properties of the filaments they contain.

The main results regarding these filaments are the following:
\begin{enumerate}
\item Given the very coherent filament detection criteria between the two extraction methods (and despite the technical disparities) the statistical properties of the detected structures in any skeleton can be considered as quasi-identical.

\item The filaments display a typical width of $\sim0.12\pc$ and have a narrow spread of $\pm0.04\pc$ (Fig.~\ref{fig:width}). This value seems to be free from detection bias and is supported by the width of the variations in the velocity field (Figs. \ref{fig:depletion} and \ref{fig:cenprof});

\item The upper ends of the distributions of linear densities and volume densities of the filaments are consistent with observations and simulations of interstellar filaments. However, many extracted filaments have lower densities. This suggests that the criteria generally used to identify filaments are too restrictive.%, but they also sport a large excess towards the lower densities. This is clearly a detection effect resulting from the very lax definition adopted for the filaments, but, as there seem to be a continuity in the sample of filaments identified in this work, one can argue that the criteria generally used to identify filaments are too restrictive, biasing the samples towards high densities only;

\item The filaments in Orion\,B are stable against gravitational collapse because of their relatively lukewarm temperatures and their moderately supersonic velocity dispersions. This is consistent with the very low SFE of Orion\,B, at about 1\%;

\item At least one filament (the Hummingbird filament, lying north of NGC\,2024) shows visible signs of periodic longitudinal fragmentation, despite being clearly gravitationally sub-critical.

\item In the vicinity of the filaments, the velocity dispersion is larger in $^{13}$CO$\,(J=1-0)$ than in \COline. The surroundings of the filaments are thus more turbulent than their inner part. In addition to that, the \COline\ velocity dispersion decreases significantly towards the centre of the filaments, almost reaching transonic values. This suggests that turbulence is being dissipated in the filaments although they are not gravitationally bound.

%\item It is difficult to clearly check whether the relative orientation of the filaments and the magnetic field differs for sub-critical and trans-critical filaments, because the resolution of the \emph{Planck} data is one order of magnitude coarser than the measured width of the Orion\,B filaments. We here tentatively favour a perpendicular orientation combined with a significant inclination of the filaments.

\end{enumerate}
These results speak in favour of the filamentary structures in Orion\,B being in a young evolutionary stage, meaning that the cloud might eventually evolve into a more active environment like its neighbour Orion\,A. This is consistent with the relatively young age of the \Hii\ regions within the cloud, estimated at $\sim 100\,000$ and $200\,000$ years for NGC\,2023 and NGC\,2024, respectively (Tremblin 2014, priv. comm.), and also with the fact that the fraction of very young protostars among all YSOs is significantly larger in the south-western part of Orion\,B than in Orion\,A \cite[24\% vs. 1.5\%,][]{stutz13}\\

This paper mostly focussed on the transverse properties of the filaments. The longitudinal properties were preliminarily tackled in the specific case of the Hummingbird filament, but these properties deserve an in-depth study as well, because signs of longitudinal fragmentation are suspected at other locations in the cloud and are apparently accompanied by longitudinal velocity patterns. The Hummingbird filament itself will be the object of a follow-up study at higher angular resolution, which will shed light onto the entire filamentary network of Orion\,B.

From a methodological point of view, the Hessian approach (Appendix \ref{app:extract}) can be easily extended to perform the filament detection in three dimensions. Such a three-dimensional detection making full use of the velocity information would facilitate the analysis of the longitudinal velocity structure of the filaments, and possibly enable the detection of fibres \citep{hacar13}.\\

Thanks to the multi-tracer nature of the ORION-B project, the filamentary skeletons could be used to stack the observed spectra to reveal faint molecular lines and characterise the molecular signature of filamentary regions, in a manner similar to \citet{gratier17} and \citet{bron18}. 
Conversely, classification tools such as the clustering used by \citet{bron18} could possibly reveal the presence of different families of filaments, based on their properties derived in this paper, which could point to different evolutionary stages or scenarios for the filaments in Orion\,B.

\begin{acknowledgements}
We thank Pr. Maciej Orkisz for pointing us to the work of \citet{frangi98} and helping us with the understanding and implementation in an astrophysical context of this analysis method coming from the field of medical imaging. We also thank Pr. Jouni Kainulainen for stimulating discussions and helpful remarks. This research has made use of data from the Herschel Gould Belt Survey (HGBS) project (\url{http://gouldbelt-herschel.cea.fr}). The HGBS is a Herschel Key Programme jointly carried out by SPIRE Specialist Astronomy Group 3 (SAG 3), scientists of several institutes in the PACS Consortium (CEA Saclay, INAF-IFSI Rome and INAF-Arcetri, KU Leuven, MPIA Heidelberg), and scientists of the Herschel Science Center (HSC). This work was supported by the CNRS/CNES programme ``Physique et Chimie du Milieu Interstellaire'' (PCMI). We thank the CIAS for its hospitality during the three workshops devoted to this project. NP wishes to acknowledge support from STFC under grant number ST/N000706/1. PG thanks ERC starting grant (3DICE, grant agreement 336474) for funding during this work. JRG thanks the Spanish MINECO for funding support under grant AYA2017-85111-P.
\end{acknowledgements}

\bibliographystyle{aa} %
\bibliography{filaments} %

\begin{appendix}
\section{Multi-Gaussian fitting of the molecular spectra}
\label{app:gauss}

In order to transform the noisy observational data into a model cube, we first perform a Gaussian smoothing the \COline{} VESPA spectra with a width of 3 velocity channels to improve the peak detection. Significant peaks in the spectrum, above a threshold of $4\sigma$, are detected, and then a multi-Gaussian fit is performed on the spectrum with one Gaussian component per identified peak. After that, the reduced $\chi^2$ is computed over 6 channels on either side of the peak (13 channels in total), and compared to the noise level of the spectrum. 

\begin{figure}
    \centering
    \includegraphics[width=\linewidth]{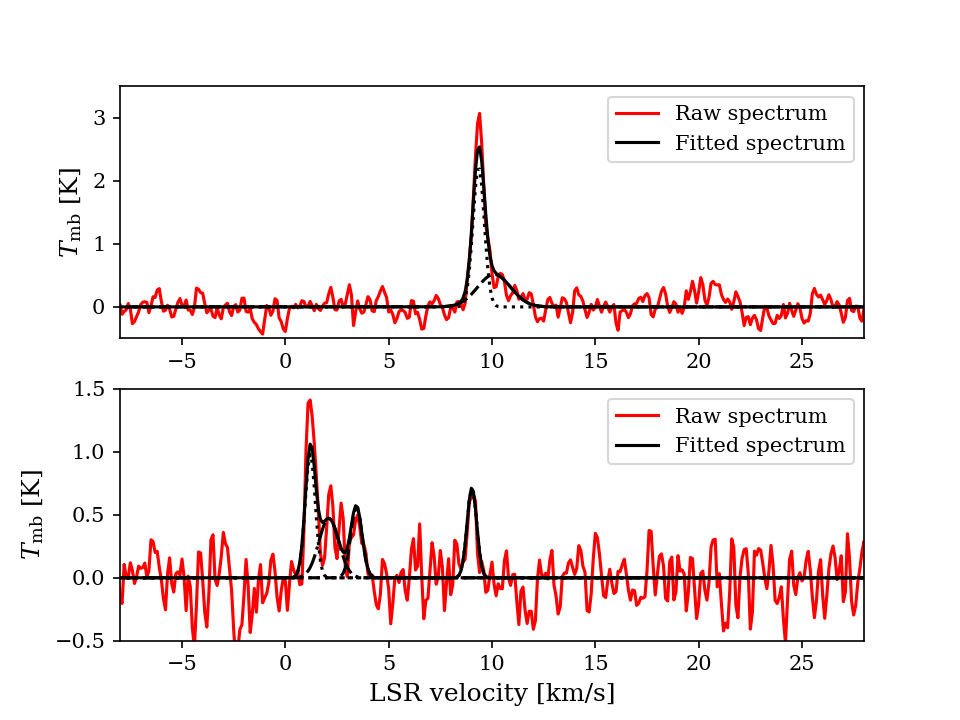}
    \caption{Examples of spectra fitted with multiple Gaussians, at signal-to-noise levels of about 15 (\emph{top}) and 5 (\emph{bottom}). The individual Gaussian components are indicated in dashed, dotted, or dash-dotted lines.}
    \label{fig:spectra}
  \end{figure}

If the $\chi^2$ is lower than $0.5\sigma$, an unnecessary Gaussian component is removed. When it is larger than a threshold of $2\sigma$, an extra Gaussian component is added in the corresponding velocity range. Our initial determination of the number of significant peaks tends to be overestimated. For that reason, during the iterative process, only one extra component can be added. Conversely, the number of components can decrease to a minimum of one. This allows the fitting process to take into account, for example components separated by less than their average velocity dispersion, which do not present one separate peak per component. But it also prevents over-fitting, whereby a single spectral component would be reproduced by many Gaussian peaks. Finally, we visually inspect the residual cube to check that no obvious signal feature remains unfitted. Figure \ref{fig:spectra} shows two examples of the fit of multicomponent spectra.

Spatial correlations between neighbouring pixels are not taken into account during the fit. However, the retrieved moment maps, and in particular the integrated intensity, do not show any significant incoherence (Fig. \ref{fig:data}, left).

\section{Structural detection of the filaments: Methodological details}
\label{app:method}
\subsection{Step by step extraction of the filaments}
\label{app:extract}

After obtaining from the raw observational data a modelled cube and a column density estimate (Fig.~\ref{fig:filS1}, panels 1, 2 and 3), we use this column density map to identify the filamentary structures in Orion\,B. This section details the extraction of the skeletons used throughout this paper. We restrict ourselves to study filamentary structures in a two-dimensional map but the concepts and implementations can be generalised to a PPV cube.

\subsubsection{Hessian approach}
We have qualitatively and observationally defined filaments as elongated, over-dense structures in the molecular ISM. Their identification can be compared with that of mountain ranges in an altitude map. These mountain ranges and their ridge lines can be simply defined in terms of topography, that is in terms of differential properties of the studied map.

The Hessian matrix (i.e. the matrix of the second order partial derivatives) provides the most useful insight into the topography: it measures the variations of the slopes, and therefore enables us to locate the valleys, summits, and ridges. The eigenvectors of the Hessian indicate the main directions of variations of the slope, and the eigenvalues ($\varepsilon_1$ and $\varepsilon_2$ in our two-dimensional case) whether the terrain is passing through a minimum or a maximum in these two directions. In the case of filaments or ridges, we expect the altitude to vary very little in one direction, and to reach a local maximum in the other direction. We therefore expect a second derivative with a value close to zero along the ridge and a strongly negative derivative perpendicular to the ridge; if we were working with absorption data, the reasoning would remain the same, albeit with reversed signs. We can summarise the behaviours of $\varepsilon_1$ and $\varepsilon_2$ in terms of characteristic structures in Table~\ref{tab:hess}.

\begin{table}
    \centering
    \caption{Topography of a two-dimensional dataset depending on the eigenvalues of its Hessian matrix $\varepsilon_1$ and $\varepsilon_2$. The parameter L indicates a low value; H- and H+ indicate a highly negative (respectively positive) value.}
    \begin{tabular}{ccc}
      \hline
      \hline
          $\varepsilon_1$ &  $\varepsilon_2$ & Structure \\
          \hline
          L & L & No preferred direction \\
          L & H- & Ridge \\
          L & H+ & Valley \\
          H- & H- & Peak (local maximum) \\
          H- & H+ & Pass (saddle point) \\
          H+ & H+ & Hole (local minimum) \\
          \hline
        \end{tabular}
    \label{tab:hess}
  \end{table}

%The Hessian is computed for the map of integrated intensity $W_0 = \!\int\!I(v) \mathrm{d}v$, which takes into account all the emission for a given line of sight, while allowing us to work on a two-dimensional field. Had we not been using a noiseless data-cube, we should have better used a map of peak temperature $T_\mathrm{peak}$ (i.e., the maximum intensity in main-beam temperature on the spectrum for each line of sight), which would have still allowed us to identify the brightest regions, without the issue of performing noisy integration along the velocity axis.

The computation of the Hessian itself makes use of the concepts of linear scale space theory \citep{florack92,koenderink84}, as advised by \citet{frangi98}. Differentiation is thus performed by convolving the field with derivatives of $n$-dimensional Gaussians, which allows us to smooth out simultaneously the noise and adapt to the typical spatial scale of searched structures \citep{frangi98,salji15}; in our case, as no initial assumption is made on the length of filaments, the only relevant scale is their width. 
Once the smooth Hessian is obtained and diagonalised, we can use the eigenvalues as a filter to extract filamentary regions from the map. In this work two implementations of such filters are used.

\subsubsection{Filament extraction by thresholding}

\begin{figure*}
    \centering
    \includegraphics[height=0.28\textheight]{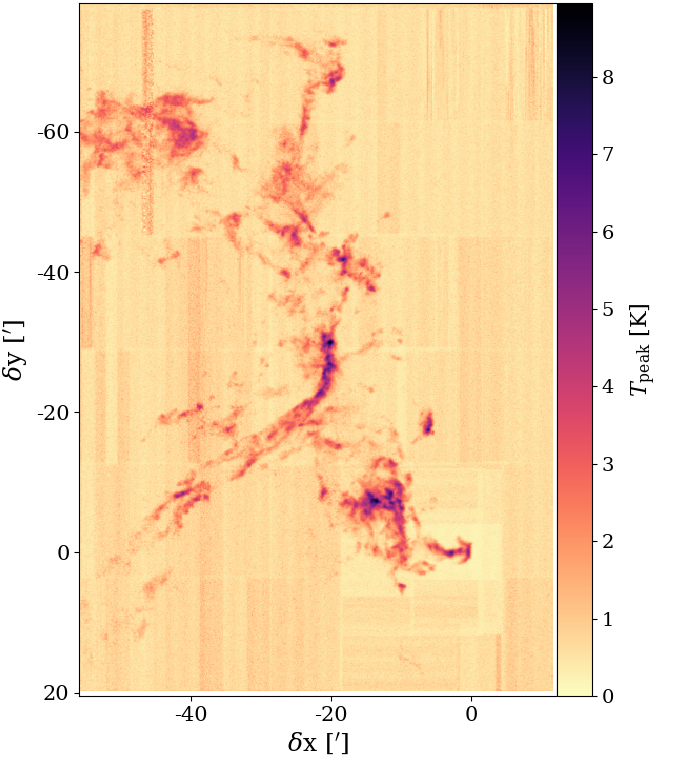}
    \includegraphics[height=0.28\textheight]{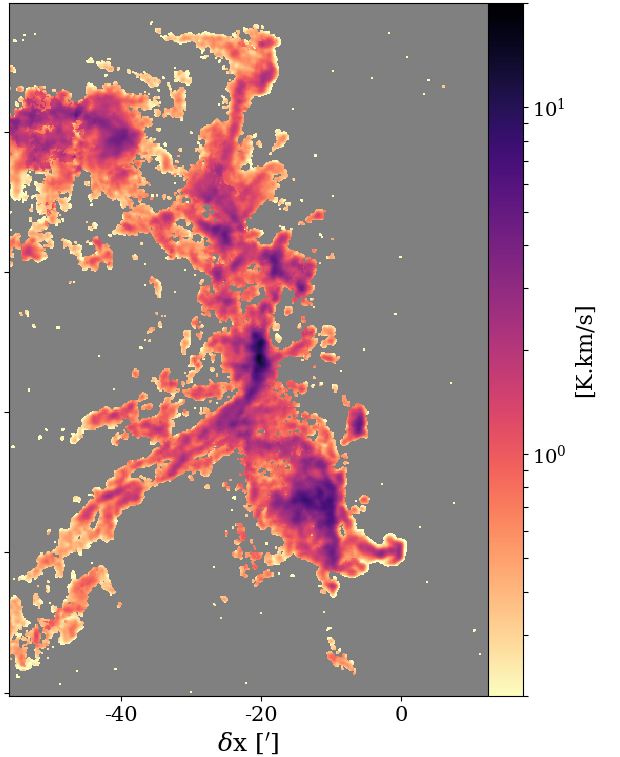}
    \includegraphics[height=0.28\textheight]{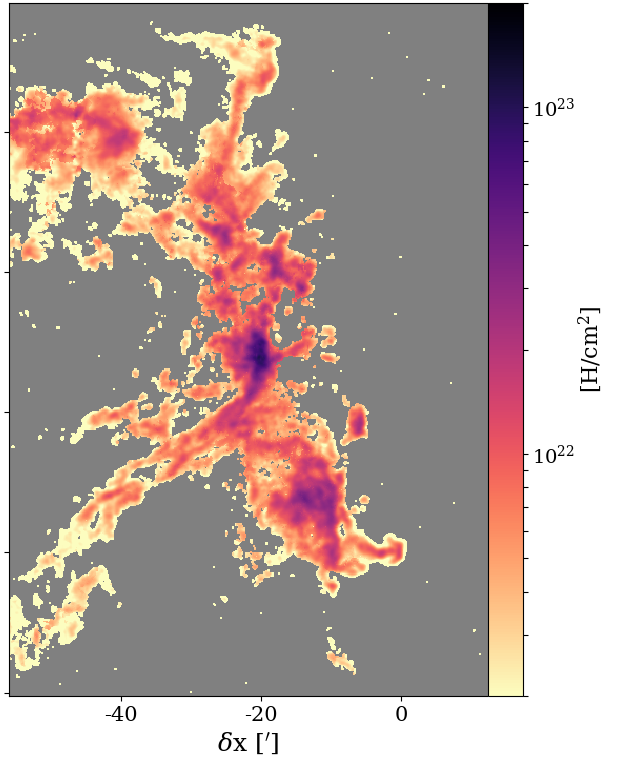}

    \includegraphics[height=0.28\textheight]{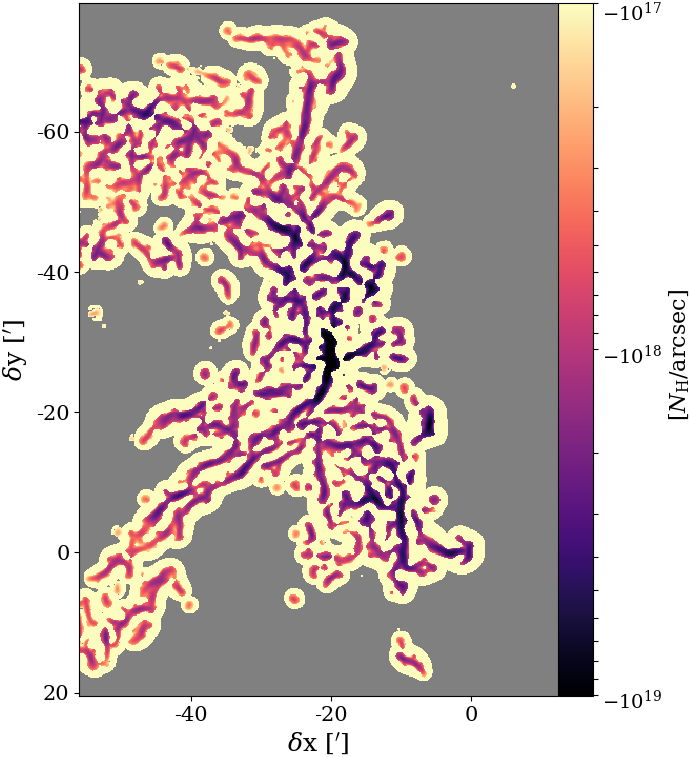}
    \includegraphics[height=0.28\textheight]{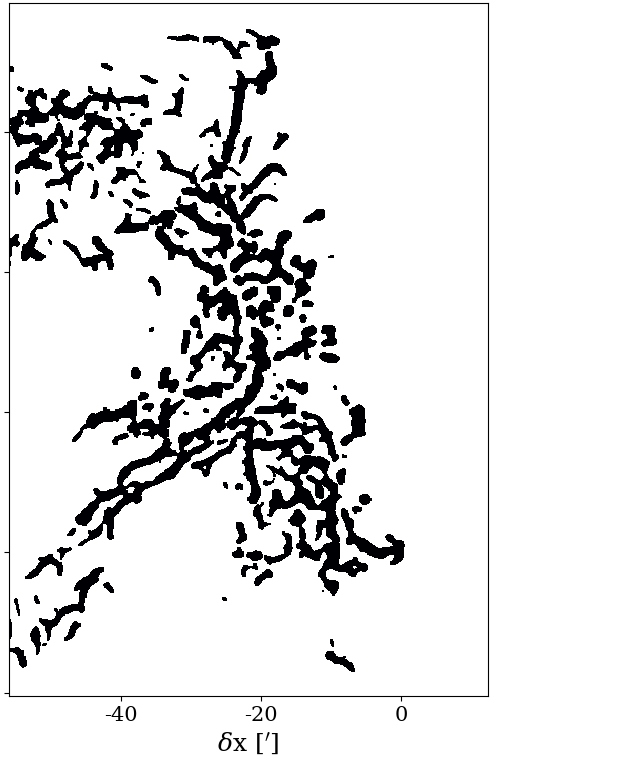}
    \includegraphics[height=0.28\textheight]{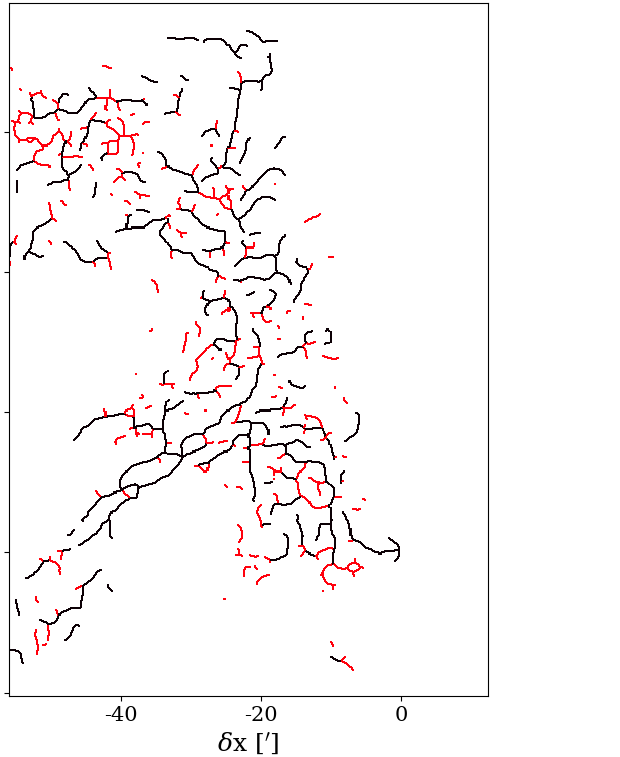}
    \caption{\emph{From left to right and top to bottom:} (1) Peak temperature of the raw \COline{} data; (2) integrated intensity of the multi-Gaussian fit result; (3) column density estimate derived from this integrated intensity and the dust temperature; (4) largest (in absolute value) eigenvalue $\varepsilon_2$ of the Hessian matrix computed at each point of the column density map; (5) regions identified as filamentary, by applying a threshold onto the $\varepsilon_2$ map; and (6) skeleton obtained by morphological thinning of these regions. The filaments that are eliminated at some point in the cleaning process are indicated in red.}
    \label{fig:filS1}
  \end{figure*}

The first approach is meant to be as simple as possible, and is similar to that presented in \citet{PCi32}. The Hessian is directly computed for the map of \CO{}-derived column density (Fig.~\ref{fig:filS1}, panel 4), and the analysis focussed on a single spatial scale set to 0.13\pc{} (corresponding to 0.11\pc{} after deconvolution). This choice of scale is the result of an iterative approach. The first guess for the size of the filaments was assumed to be 0.1\pc{}, following \citet{arzoumanian11}. The distribution of filament widths yielded a peak consistent with this initial guess, which could have been a detection bias. The analysis was therefore repeated with a Hessian smoothing scale of 0.05\pc{} and 0.2\pc{}, and in both cases the peak of the filament width distribution was of 0.11$\pm0.01\pc$. Therefore the final choice of detection scale is set to match as well as possible the scale properties of the dataset.

We then threshold the eigenvalues. This is a quantitative way to transcribe the characteristics of a filament in terms of eigenvalues, as seen in Table~\ref{tab:hess}. If the eigenvalues are first sorted so that $|\varepsilon_1| < |\varepsilon_2|$, then selecting the pixels where $\varepsilon_2 < \tau < 0$ ($\tau$ being a global threshold) ensures that we are in the vicinity of a local maximum in the direction perpendicular to the filament. No condition is set on the smaller eigenvalue, so that we allow any kind of peak or saddle point, as long as the relief is not steeper along the ridge than perpendicular to the ridge (Fig.~\ref{fig:filS1}, panel 5).
We find the best value of the threshold $\tau$ to be 4\% of the lowest (negative) eigenvalue in the map, based on visual inspection (Fig.~\ref{fig:filS1} left).

\subsubsection{Adaptive filament extraction with ridge detection}
\label{sec:S2}
\begin{figure*}
    \centering
    \includegraphics[height=0.28\textheight]{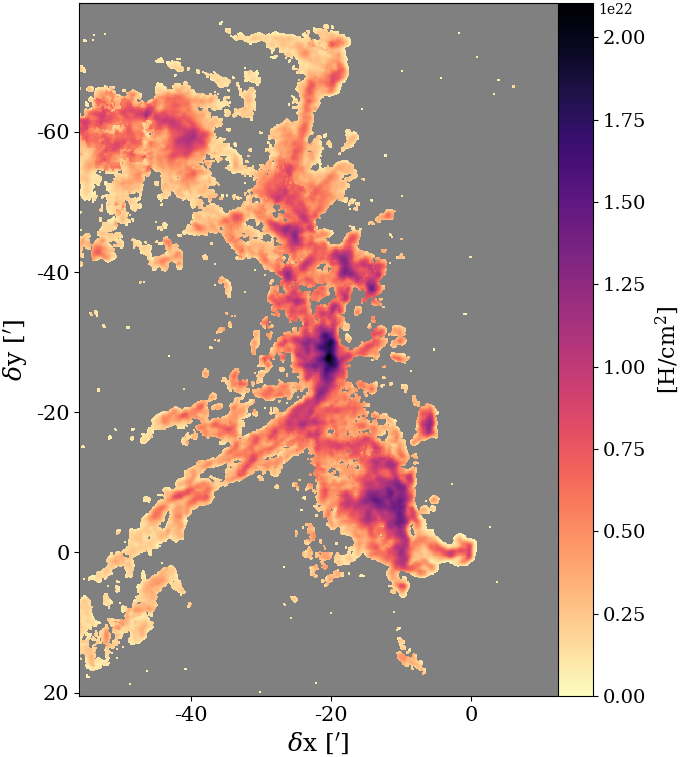}
    \includegraphics[height=0.28\textheight]{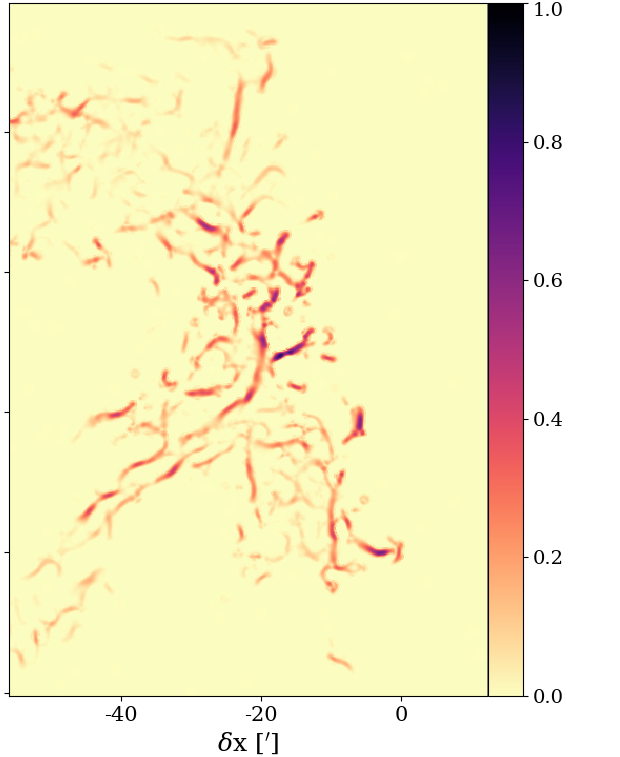}
    \includegraphics[height=0.28\textheight]{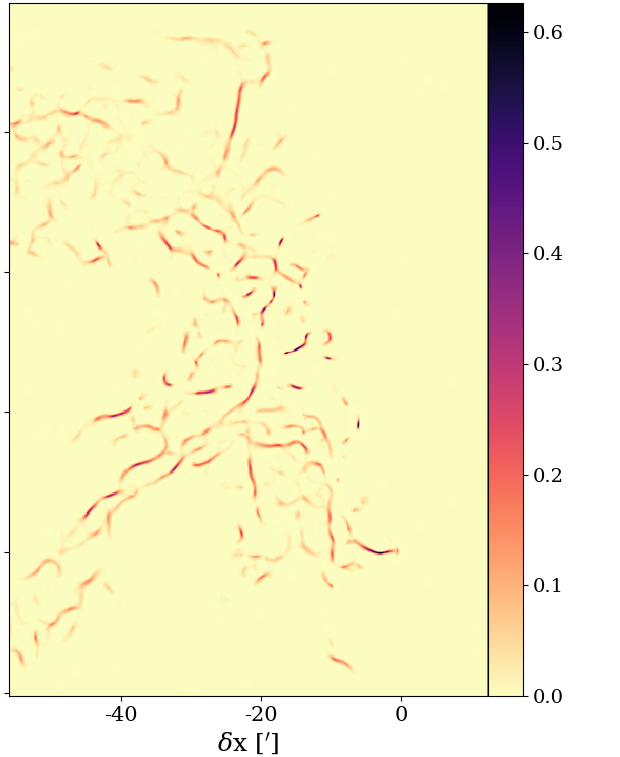}

    \includegraphics[height=0.28\textheight]{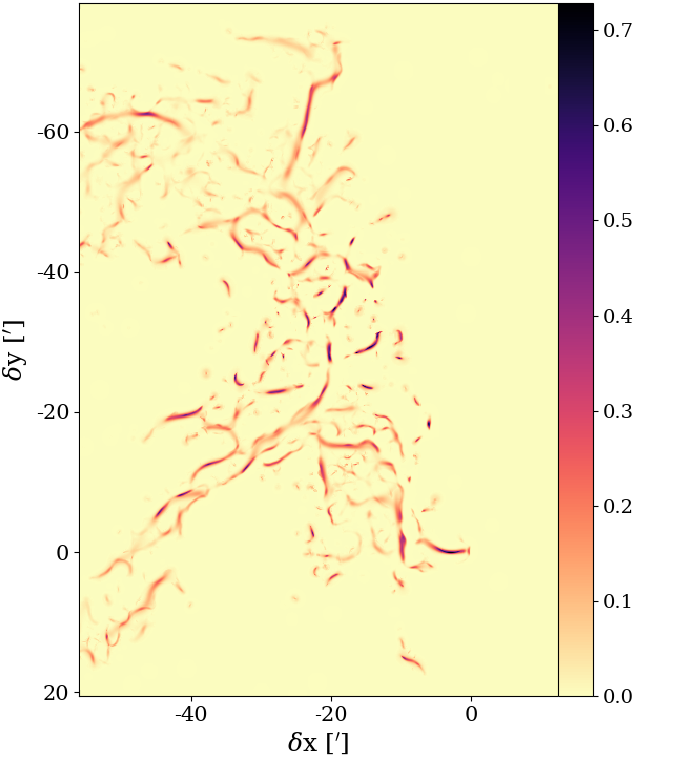}
    \includegraphics[height=0.28\textheight]{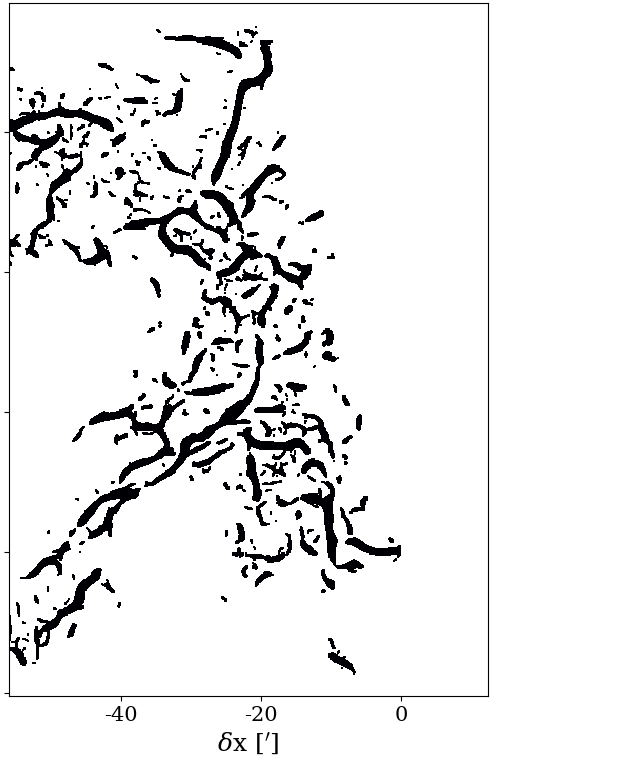}
    \includegraphics[height=0.28\textheight]{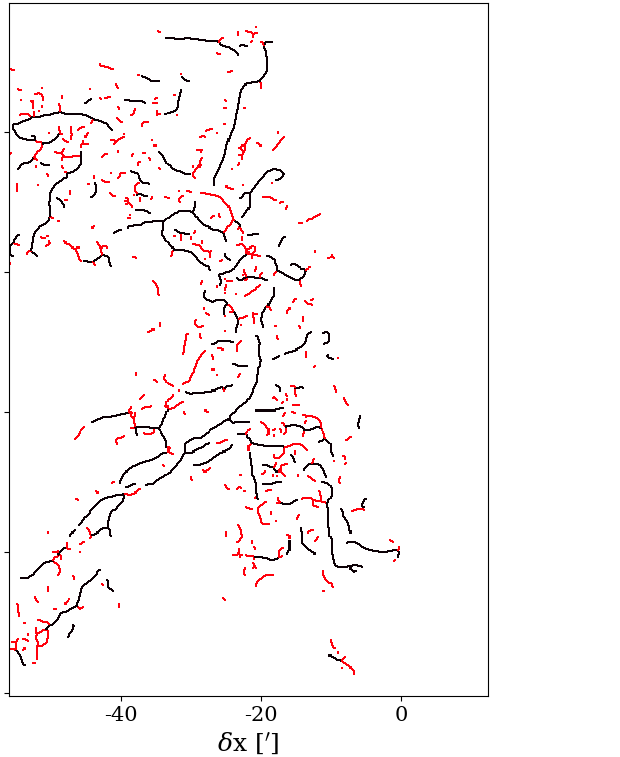}
    \caption{\emph{From left to right and top to bottom:} (1) Column density (Fig.~\ref{fig:filS1}, panel 3) after an asinh transform (colour scale is linear, not logarithmic); (2) structures resulting from the aspect-ratio filter of \citet{frangi98} applied to the Hessian eigenvalues of the transformed column density, with a single detection scale of 0.14\,pc; (3) the same structures after applying a gradient-based ridge-detection filter, with the same detection scale; (4) the result of the multi-scale (0.06 to 0.3\pc) filtering; (5) regions identified as filamentary, by applying a threshold onto the multi-scale filtered map; and (6) skeleton obtained by morphological thinning of these regions. The filaments that end up being eliminated at some point in the cleaning process are indicated in red.}
    \label{fig:filS2}
  \end{figure*}

For the second approach, several additional features refine the filament detection method.
\citet{koch15} applied an arctan transform to the intensities to suppress bright point sources (e.g. dense cores) that can dominate the fainter filamentary structures being are searched for. %The column density of a GMC is composed of three main  components: 1) An extended component (diffuse and translucent gas) that is filtered out in our case as this medium is not dense enough to brightly emit in \CO{}; 2) A network of intermediate density filaments; and 3) bright dense cores. The largest column density pixels in our map belong to dense cores that appear as very bright, almost point sources. 
In our case, we find it better to use the asinh transform used in \citet{gratier17} (Fig.~\ref{fig:filS2}, panel 1), which is linear at small intensities and logarithmic at high intensities. The asymptotic behaviour of the arctan transform would flatten the bright regions and
thus render the Hessian approach ineffective. We therefore use
\begin{equation}
\widetilde{\Nh} = a\cdot \mathrm{asinh}(\Nh/a)
\end{equation}
as an input for filament detection with the parameter $a$ set to $a = 5.18\times10^{21}\pscm$, i.e. the median of non-zero values of the \CO{}-derived column density.

This second filament extraction method takes into account the local aspect ratio of the studied field. The concepts behind this method were first described in the field of medical imaging for the purpose of identifying blood vessels in magnetic resonance imaging (MRI) or computer tomography (CT) images \citep{frangi98}. These concepts have been adapted to astrophysical data and extensively illustrated by \citet{salji15}. The map of Hessian eigenvalues computed from $\widetilde{\Nh}$ is filtered with the following function:

\begin{equation}
V_0 = \begin{cases}
0 & \mathrm{if~} \varepsilon_2 > 0,\\
\exp\left(-\frac{R^2}{2b^2}\right)\left(1-\exp\left(-\frac{S^2}{2c^2}\right)\right) & \mathrm{otherwise,}
\end{cases}
\end{equation}
where $R = \varepsilon_1 / \varepsilon_2$ is the local aspect ratio (here again $|\varepsilon_1| < |\varepsilon_2|$), and $S = \sqrt{\varepsilon_1^{\phantom0 2}+\varepsilon_2^{\phantom0 2}}$ is the local amplitude of the second derivative. In other terms, the filter function emphasises pixels where the signal varies significantly ($S > c$), and the local aspect ratio is large ($R < b$).

Following the recommendations of \citet{frangi98}, the parameter $b$ was set to $b=0.5$, and $c$ was set to half the maximum value of the Hessian norm $S$ over the field of view. The resulting filtered map can be seen on Fig.~\ref{fig:filS2} (panel 2).

An extra step is added by implementing a ridge-detection function, which is, as far as we know, a novel addition, but can be compared to the centreline extraction method of \citet{aylward02}. The goal  is to narrow down the response of the filter, so that the obtained structures are as close as possible to the ridge lines (i.e. the skeleton) of the filaments, rather than being broad filamentary regions. This function makes use of the fact that the component of the gradient perpendicular to the filament should go through zero at the ridge line, hence the norm of the gradient should reach a local minimum also close to zero if the variations along the ridge line are negligible. We therefore compute the norm of the gradient of our map, $|\vec{g}|=|\vec{\nabla}\widetilde{\Nh}|$, again through convolution with Gaussian derivatives. The final filter is

\begin{equation}
V = V_0\cdot\exp\left(\frac{-|\vec{g}|^2}{2d^2}\right)
,\end{equation}
where the parameter $d$ ensures that the slope $|\vec{g}|$ is close enough to zero. This parameter was set to 10\% of the maximum value of $|\vec{g}|$ over the field of view (Fig.~\ref{fig:filS2}, panel 3).

To avoid favouring a single scale, we perform this structure extraction in a multi-scale fashion, as advised by \citet{frangi98}. We run the detection algorithm to obtain the filtered map $V(s)$ with smoothing scales $s$ from 0.06\pc{} to 0.32\pc{} in steps of 0.02\pc{}, and for each pixel we pick up the maximal response among these filters to achieve a scale-adapted filament detection. Thus, the wider structures are detected strongly and stand out well, while the narrower objects in the field are still picked up (Fig.~\ref{fig:filS2}, panel 4). Once again the choice of scales is the result of an iterative approach. A first computation was performed with a 0.1\pc{} scale only, yielding filaments widths ranging from 0.06\pc{} (our resolution limit) to about 0.3\pc{}. Although the widest filaments were rather poorly identified, we kept a 0.06 -- 0.32\pc{} detection range to avoid as much as possible a detection bias, while keeping the scale range reasonably limited.

After that, a global threshold $\tau_\mathrm{f}$ is applied on the resulting scale-adapted filtered map (which now has values ranging from 0 to 1) to select regions which are close enough to the ridge lines of filaments. By visual inspection, $\tau_\mathrm{f}$ was set to 0.03 (Fig.~\ref{fig:filS2}, panel 5).

\subsubsection{Skeletonisation}
Once binary masks that identify the filamentary regions have been obtained, they are thinned down until we are left with single-pixel wide skeletons. One of the most common methods of skeletonisation is known as morphological thinning, which we perform following the algorithm of \citet{zhang84}, as implemented in the Python module \texttt{skimage}. The resulting skeletons are shown in the last panels of Fig.~\ref{fig:filS1} and \ref{fig:filS2}. As this step of the analysis only makes use of the geometry of the binary masks and does not take into account the underlying physical map, the extracted centrelines do not necessarily match the ridge lines of the filaments, especially if they do not have a cylindrical geometry. It is therefore all the more useful to try and narrow down the mask as accurately as possible before the skeletonisation, as done with the gradient filter in Sect.~\ref{sec:S2}.

\subsection{Cleaning the skeleton}
\label{app:clean}

Separating a filamentary skeleton into individual filaments enables us to clean it by checking if each individual filament matches the assumed definition. For that purpose, a geometrical analysis allows us to distinguish between regular points and nodes (or vertices). The regular points have exactly two neighbours, while the nodes have fewer (if they are endpoints) or more (if they are intersections) than two. The individual filaments are therefore strings of points belonging to the skeleton and linking two nodes.
The cleaning process relies on the following criteria: skeleton geometry, curvature radius, relative contrast, and width of the individual filaments.

\subsubsection{Geometrical cleaning}

The first stage of geometrical cleaning is applied immediately after the first extraction of the filamentary skeleton. It consists in removing isolated nodes (i.e. single pixels) and very short filaments: to be able to distinguish between clumps and filaments, we require an aspect ratio of at least 2 and therefore we set a lower length limit at 0.22\pc, i.e. twice the typical filament width. As a dense skeleton can contain many short individual filaments owing to the frequency of the intersections, the length selection criterion is only applied to individual filaments that are either isolated or that are so-called spurious branches; i.e. filaments that stick out off the side of a structure and for which only one of their extremity is an intersection as the other is an endpoint.

This stage of cleaning is also reapplied after each of the other stages of cleaning, usually twice in a row to take into account the structure differences that can result from the elimination of spurious branches. Experience shows that two repetitions of this geometrical cleaning are enough to converge to a stable cleaned skeleton; the second repetition is even often superfluous.

\subsubsection{Curvature radius}
\label{app:curv}

One of the first possible characterisations of the individual filaments is a quantitative analysis of their shape, namely whether the individual filaments are rather straight or very curvy. Several methods are possible to measure how straight or curvy a filament is. A simple approach is the measurement of the ratio of the distance between the extremities of the filament and its curvilinear length, but this does not discriminate between a large smooth loop and a jagged structure that is more likely to be an artefact of skeletonisation. Another approach is to compute the circular variance of the filament position angles. But again a large smooth loop might be discarded as its position angles are spread around 360\degr{}, while a poorly identified filament with a position angle varying wildly from $-70$ to $+70\degr$ might have a lower circular variance and could be retained. The chosen approach was thus to use the mean curvature radius of the filament, which would discard structures with rapid position angle variation, while keeping large-scale loops.

To that purpose, the first step is estimating the position angle of the filaments at each position. Two possible approaches are possible: either the position angle is directly deduced from the eigenvectors output by the Hessian matrix diagonalisation, or it can be geometrically computed on a pixel-by-pixel basis, by comparing the position of each point in the filaments with its nearest neighbours. The first option has the advantage of yielding continuous angles, whereas the second  yields angles in steps of 22.5\degr{}.  We use the Hessian angle in the entire paper, and in
particular for curvature measurements, because this first approach is reasonably reliable despite some misalignments of the skeletons and position angle map.

The curvature radius is then simply defined as the inverse of the derivative of the position angle computed along the filament. This curvature radius $R_\mathrm{c}$ is then averaged over the individual filament, and is compared to the width $w$ of the individual filament (Sect.~\ref{sec:width}). The individual filaments are rejected if they have a $R_\mathrm{c}/w$ ratio lower than 1.5, thus ensuring that the transverse profiles of the filament are not contaminated by a further part of the filament after a sharp turn. This selection affects a large number of individual filaments, mostly short ones, as can be see in Table~\ref{tab:cleaning}.

\subsubsection{Contrast}
\label{app:contrast}

The value of the baseline at the peak of the filament profile, obtained from the Gaussian fit (Eq. \ref{eq:profile}), gives us an estimate of the surface density of the underlying background. An instinctive way to estimate the ``contrast'' of the filaments, i.e. how much they stand out from the background, is to take the ratio of the linear density of the filament to the surface density of the background for each line of sight. The resulting quantity has the dimension of a length, and can be interpreted as the distance from the filament needed to accumulate as much mass from the surrounding medium as there is in the filament. We can make this contrast dimensionless by defining the ``relative contrast'', which is the ratio of the contrast to the filament width. To summarise,

\begin{equation}
\begin{split}
\text{Contrast} &= \frac{\text{Filament linear density}}{\text{Background surface density}} \\
\text{Relative contrast} &= \frac{\text{Contrast}}{\text{Filament width}}
\end{split}
.\end{equation}

The relative contrast measures how many times denser the filament is relative to its surrounding medium: for instance, a relative contrast of ten indicates that a region about ten times broader than the filament is needed to accumulate as much mass as contained in the corresponding portion of filament. This accumulation of mass might be interesting when studying the power-law-like extension of the profile of an isolated filament, but in skeletons as crowded as those studied in this paper, we are mostly concerned by the presence of a clearly visible inner part of the filaments -- hence a lower limit imposed on the relative contrast at $\sim 1$. The distribution of relative contrasts in the non-cleaned filamentary skeletons featured a gap at a relative ratio of 0.6, which we thus chose as a limit under which filaments were rejected. As can be seen in Table~\ref{tab:cleaning}, this criterion only had a moderate impact on the cleaning of the skeleton.

\subsubsection{Final selection and summary}

In addition to the selection criteria described above, the widths of the filaments are also taken into account to reject filaments that we deem not reliably detected. Two limits are set. First, the lower limit at 0.06\pc, which{} corresponds to the resolution of the data; structures of this width or less are unresolved and thus cannot be characterised reliably.\ Second, an upper limit of 0.22\pc{}, which corresponds to a gap in the filament width distribution. The few wider individual filaments are considered as outliers. The lower limit mostly affects the smallest scale structures detected in the S2 skeleton. These narrow rejected objects are numerous. Indeed, the Hessian filter is a derivative, which by essence enhances small-scale, noisy structures, and when the Gaussian smoothing scale comes down to the 0.06\pc{} spatial resolution of the data, it stops preventing this noise enhancement, and thus many noisy structures at very low column densities are picked up by the filter. The upper limit, on the other hand, mostly affects ``bridges'' between two neighbouring filaments, which are artefacts of the skeletonisation process and are perpendicular to the orientation of well-detected filaments running close to each other.

\begin{table}
    \centering
    \caption{Evolution of the skeletons S1 and S2 after successive stages of cleaning in terms of number of pixels (Pix) and filaments (Fil). The initial skeleton is first geometrically cleaned of short, isolated filaments, then we exclude filaments with too strong curvatures, too low relative contrasts, too broad or too narrow column density profiles. See text for details on why the numbers do not add up properly.}
    
    \begin{tabular}{ccccc}
      \hline
      \hline
       Skeleton & \multicolumn{2}{c}{S1} & \multicolumn{2}{c}{S2}  \\
      \hline
       & Pix & Fil & Pix & Fil \\
      \hline
      Raw state & 5256 & 457 & 4837 & 447 \\
      Geometry cleaning & -646 & N/A & -909 & N/A \\
      Initial state & 4610 & 291 & 3928 & 246 \\
      Curvature cleaning  & -814 & -88 & -924 & -96 \\
      Width cleaning  & -472 & -43 & -675 & -67 \\
      Intermediate state & 3664 & 157 & 2833 & 119 \\
      Contrast cleaning  & -325 & -20 & -112 & -9 \\
      Final state & 3315 & 130 & 2708 & 106 \\
      \hline
    \end{tabular}
    \label{tab:cleaning}
  \end{table}

Table~\ref{tab:cleaning} summarises all stages of the cleaning process. After the first detection step, the geometrical cleaning is run twice and the resulting skeletons are used as a starting point (initial state) for the first analysis. This step starts by measuring the width and curvature of the filaments, leading to the rejection of a large amount of structures, some of which can be rejected by more than one criterion. After this, the geometrical cleaning is reapplied twice, yielding intermediate skeletons which then undergo the contrast cleaning. This is the last stage of cleaning  because of the relative contrast computation is very sensitive to the quality of the transverse profile fitting, and therefore can lead to results too extreme to be analysed for some of the poorly defined structures that get eliminated in the previous stages of cleaning. After this, two last passes of geometrical cleaning yield the final skeletons, which are then used for the rest of the analysis in this paper. We note that the numbers of pixels and filaments in Table~\ref{tab:cleaning} do not always add up properly. This is because some structures can be eliminated based on several criteria and because of geometrical cleaning, which removes some isolated nodes left out by other cleaning stage and  can reduce the number of filaments either by removing the smallest or by merging these into larger filements when a spurious branch (and thus an intersection) is eliminated.

\subsection{Qualities and limitations of the filament detection methods}
\label{app:skel-quality}

Having used in parallel skeletons obtained with two different methods, we need to compare the relative merits of each with respect to the implementation, detection quality, and physical implications.

The first method (which yields the skeleton S1) presents the major advantage of being very straightforward in its implementation, providing a simple yet efficient way to identify filamentary regions in a molecular cloud. However, it is not very specific in its geometric requirements when identifying filaments. This means that regions with strong spatial features that are not filamentary are detected, only to be later rejected during the cleaning process (see the filaments indicated in red in Fig.~\ref{fig:filS1}, panel 6), and that the identified filamentary regions are broad. This broadness is an advantage in terms of completeness, for example if we want to use the mask of Fig.~\ref{fig:filS1} (panel 5) to obtain molecular line ratios inside and outside the filaments, but it is a disadvantage during the skeletonisation, as the position of the skeleton is less precise. This uncertainty on the position of the skeleton can be a problem when measuring the properties of the filament profiles, in particular their widths (Sect.~\ref{sec:width}). The binary aspect of the detection method achieved with a single threshold also makes the method unsuitable for a multi-scale approach.

The second method, which yields the skeleton S2, is much more demanding as far as  the properties of the identified structures are concerned, which also comes with its lot of pros and cons. The rescaling of the data, which could easily be transposed to the first method, makes the whole detection process less sensitive to the brightest/densest regions, allowing for a better tuning of the filtering parameters. The first stage of filtering achieves a better distinction between filament-like and blob-like structures using the local aspect ratio. Together with the second stage of filtering, it yields much thinner filamentary regions (Fig.~\ref{fig:filS2}, panel 5), enabling a more accurate skeletonisation. The use of a continuous filter (with values ranging from 0 to 1) before thresholding makes the method suitable for a multi-scale approach, which reduces possible detection biases. However, as can be seen from the filaments indicated in red in Fig.~\ref{fig:filS2} (panel 6), this approach also yields a large amount of small-scale structures that end up being rejected during the cleaning;  there is 19\% of rejection in the first geometrical cleaning stage, compared to only 12\% in the case of the S1 skeleton. The multi-stage filtering process also requires more parameters to be adjusted by hand (at least two, $d$ and $\tau_\mathrm{f}$) than the first method. Another issue arises from the smoothing that occurs during the computation of the gradient. The gradient can be distorted in the vicinity of intersections of filaments, which can result in a distorted (misaligned) skeleton, but more often the filter simply misses the intersections, thereby truncating the filaments.

In summary, we can consider to first order that S1 is an upper limit for filament detection and S2 is a lower limit. This would be completely true if S2 were nested in S1 (i.e. if we had S2~= robust~$\subset$ S1). However, Fig.~\ref{fig:filcompare} shows that this is not the case, but that nonetheless filaments exclusive to S2 are far less numerous than those exclusive to S1.

It is important to note that the final quality of the obtained skeletons does not only depend on the initial detection method, but also on the cleaning process. In that respect, we can see that a larger proportion of the S2 skeleton (44\%) is rejected than in the S1 skeleton (37\%). In both cases it is a significant fraction, which shows that the skeletons should not be used without cleaning, lest the statistical results be contaminated by many unwanted structures. We can also add a qualitative remark on the rejected fraction of the skeletons: although the fraction rejected in S1 is smaller, the filaments rejected in S2 seem to be mostly very small structures, while the longer ones are retained. It can be a sign that the multi-scale detection has a drawback tightly linked to its main quality. Because this detection is unbiased, it picks up many small features that, on closer inspection, end up not matching the requirements that we set on \emph{bona fide} filaments. It is thus not easy to rule in favour of one or the other detection scheme -- at least not in their current state.\\

We can also question the nature of the detected structures, regarding their geometry and their origin. Filaments are most often described as elongated, almost unidimensional condensations in the three-dimensional turbulent medium, but this mechansm is not the only one that can form elongated structures in the molecular ISM. Geometry and projection effects are an important issue. For example, it is possible that some detected filaments are actually two-dimensional structures seen edge-on. This is most probably the case for at least one filamentary structure detected in Orion\,B, namely the vertical filament at the base of the Horsehead Nebula. Rather than a unidimensional structure, this filament is the edge of the IC\,434 ionisation front, a wall seen edge-on. Thus, its high observed density is the result of a larger dimension along the line of sight than for other filaments. Another structure can raise the question of its evolutionary scenario: the Horsehead Nebula itself. Rather than a condensation in a three-dimensional turbulent medium, it is rather a pillar carved by the IC\,434 ionisation front.

There are two reasons to  keep all structures in the filamentary network, even when the knowledge of the region points to the fact that they might not match the usual definition of filaments. First, this a priori knowledge of the structure can be absent when studying another filamentary region, and it is important to treat the skeleton statistically in an unbiased way, to see whether and how the physical properties of the filaments can distinguish different populations. The results of this paper suggest that all the above-mentioned structures should be treated in the same way. Second, mostly in the case of the Horsehead Nebula, the fact that an elongated structure had one formation scenario or another does not necessarily determine its further evolution, hence the necessity of keeping the entire variety of structures in the sample.

%\section{Statistics of the filamentary network}
%\subsection{Length of the individual filaments}

%\subsection{Linear density}

%\subsection{Volume density and fraction of traced mass}

\end{appendix}

\end{document}